\documentclass[preprint,12pt]{elsarticle}
\usepackage{subfigure}
\usepackage{lineno,hyperref}
\usepackage{indentfirst}
\usepackage{bm}
\usepackage{threeparttable}
\usepackage{multirow}
\usepackage{mathrsfs}
\usepackage{amsmath}
\usepackage{array}
\usepackage{amssymb}
\usepackage{graphicx}
\usepackage{url}
\usepackage{float}
\usepackage{hyperref}
\usepackage{epstopdf}
\usepackage{booktabs}
\usepackage{color}
\usepackage{lscape}
\usepackage{fancyhdr}
\usepackage[top=2.5cm, bottom=2.5cm, left=2cm, right=2cm]{geometry}
\modulolinenumbers[5]

\journal{?}

\biboptions{numbers,sort&compress}

\begin{document}

\begin{frontmatter}

\title{The study of coherent Rayleigh-Brillouin scattering in multiple flow regimes using unified gas-kinetic scheme}

\author{Xiaozhe Xi$^a$}
\ead{xxiab@connect.ust.hk}

\author{Junzhe Cao$^a$}
\ead{jcaobb@connect.ust.hk}

\author[]{Kun Xu$^{a,}$$^{b,}$$^{c,}$$^*$\corref{mycorrespondingauthor}}
\ead{makxu@ust.hk}

\address{$^a$Department of Mathematics, Hong Kong University of Science and Technology, Hong Kong, China\\
$^b$Department of Mechanical and Aerospace Engineering, Hong Kong University of Science and Technology, Hong Kong, China\\
$^c$HKUST Shenzhen Research Institute, Shenzhen, 518057, China}

\begin{abstract}

Coherent Rayleigh–Brillouin scattering (CRBS) holds great promise for the characterization of gas properties and the investigation of gas kinetic processes. The CRBS spectrum exhibits a strong dependence on the Knudsen number (Kn), revealing its inherently multiscale nature. In the unified gas-kinetic scheme (UGKS), collisions are intrinsically coupled with free transport during flux construction, endowing the method with distinct multiscale capabilities. Specifically, the UGKS reduces to a Boltzmann solver when the relaxation time $\tau \ge \Delta t$, and to the gas-kinetic scheme (GKS)—a Navier–Stokes solver—when $\tau \ll \Delta t$, thereby accommodating flow regimes without constraints on the molecular mean free path or collision time.
In this study, the UGKS is extended to simulate CRBS phenomena, with the governing equation formulated based on the BGK–Shakhov model. Detailed derivations are provided. To account for the additional perturbation source term, a second-order accurate numerical algorithm is developed using the Strang splitting method within the UGKS framework. The proposed model is validated against argon CRBS experiments, demonstrating excellent agreement. Building on this validated framework, the impact of incident signal intensity on CRBS spectra across a range of Knudsen numbers is systematically examined, accompanied by an in-depth analysis of the underlying physical mechanisms.
This work broadens the applicability of CFD-based CRBS simulations and provides a reliable numerical foundation for exploring high-intensity, multiscale gas-kinetic phenomena in future research.

\end{abstract}


\end{frontmatter}

\section{Introduction}\label{Sec: introduction}

Due to molecular motion, light undergoes scattering as it propagates through a gas. In most cases, the resulting scattered spectrum exhibits multiple peaks, each associated with a distinct scattering mechanism\cite{young1982rayleigh, grinstead2000coherent, wu2022rarefied}. This feature is particularly evident in Rayleigh–Brillouin Scattering (RBS), whose spectrum primarily comprises two components: Rayleigh and Brillouin scattering. Rayleigh scattering is a linear elastic process arising from Doppler shifts induced by the thermal motion of individual gas molecules and exhibits a much larger intrinsic cross-section than inelastic processes such as Raman scattering. In contrast, Brillouin scattering is a collective process involving acoustic phonons within the gas. The coherent molecular motion associated with these phonons behaves as a set of “macroscopic scatterers”, concentrating the energy and photon distribution of the scattered light and thereby enhancing the effective scattering cross-section\cite{miles2001laser}.
Owing to its relatively large scattering cross-section, RBS enables the retrieval of macroscopic fluid parameters through rapid spectral measurement and fitting, making it a powerful non-destructive optical diagnostic technique for gas characterization. This method has found broad applications in meteorological observation and atmospheric remote sensing\cite{zhai2020rayleigh, pi2025rapid}, hypersonic wind tunnel testing\cite{panda2020spectrally}, plasma diagnostics\cite{shneider2007molecular, barker2001optical}, and investigations of combustion kinetics\cite{trueba2025filtered}.

\begin{figure}[H]
	\centering
    \includegraphics[width=0.5 \textwidth]{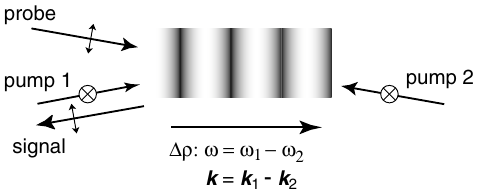}
	\caption{The physical process of CRBS in gas\cite{PhysRevLett.89.183001}.}
    \label{fig: The physical process of CRBS}
\end{figure}

Building on its application in fluid parameter measurements, researchers have developed and investigated various forms of Rayleigh–Brillouin Scattering (RBS), notably spontaneous RBS (SRBS) and coherent RBS (CRBS). In SRBS, incident light with a given wave vector is scattered due to spontaneous fluctuations in gas density\cite{srbswangphd, park2025physics, zhang2009monte, ma2021molecular}. Conversely, CRBS operates through the generation of periodic refractive-index gratings within a medium. These gratings are produced by the interference of two pump laser beams, which form a spatially periodic light-intensity pattern inside the medium. The physical mechanism of CRBS in gases is illustrated in Fig.\ref{fig: The physical process of CRBS}.
By linearly varying the frequency difference between the two pump beams, interference patterns with different velocities can be created. Under the influence of the ponderomotive force, gas particles accumulate in the high-intensity regions of the interference pattern, forming an optical lattice. When a third beam—referred to as the probe beam—is incident on this lattice at an angle satisfying the Bragg condition, it undergoes coherent scattering, generating a signal beam. Measuring the signal beam’s intensity as a function of the lattice velocity yields the RBS spectrum\cite{PhysRevLett.89.183001}.
The CRBS spectrum typically exhibits a triplet structure, comprising a Rayleigh peak centered at zero frequency difference and broadened by Doppler effects due to molecular thermal motion, along with two Brillouin side peaks symmetrically positioned about the Rayleigh peak. These side peaks, centered near the medium’s sound speed, result from collective acoustic fluctuations. Unlike SRBS, where scattering occurs isotropically, CRBS confines the scattered light to a specific direction through phase matching, thereby achieving a much higher signal-to-noise ratio. In the kinetic regime, the CRBS line shape serves as a powerful diagnostic tool for probing gas properties and exploring microscopic kinetic processes.

Among the existing numerical studies of Coherent Rayleigh–Brillouin Scattering (CRBS), several kinetic line-shape models for Maxwellian molecules have been developed, such as the Tenti S6 model\cite{tenti1974kinetic} and Pan’s S6/S7 models\cite{PhysRevLett.89.183001}. These models are derived from simplified linearized Boltzmann equations—specifically, the linearized Bhatnagar–Gross–Krook (BGK) equation for atomic gases and the Wang–Chang–Uhlenbeck equation for molecular gases—and have shown good agreement with experimental observations.
Under weaker modeling assumptions, the Direct Simulation Monte Carlo (DSMC) method has also been applied to CRBS systems both with\cite{suzuki2024effects} and without\cite{cornella2012experimental} chirped lasers to simulate the rarefied flow regime. However, DSMC is computationally expensive due to inherent statistical noise, especially for unsteady problems. Moreover, because DSMC decouples molecular free transport and collisions, its spatial and temporal resolutions are constrained by the molecular mean free path and mean collision time, respectively, leading to substantially higher computational cost in the near-continuum regime.
In this work, the deterministic Unified Gas-Kinetic Scheme (UGKS)\cite{xu2010unified} is extended for CRBS simulations. By directly modeling the underlying physical laws within a discretized space, UGKS transcends the microscopic scale limitations typical of near-continuum flows. Specifically, its integral solution of the kinetic model is used to construct fluxes that couple free transport and collision effects of the gas distribution function, recovering the Gas-Kinetic Scheme (GKS)\cite{xu2001gas} in the macroscopic limit. The UGKS has been successfully applied to a range of physical systems, including hypersonic flows\cite{long_implicit_2024, jiang2019}, micro flows\cite{liuchang2020}, thermal and chemical non-equilibrium flows\cite{wei2024adaptive, wei2025unified}, radiation transport\cite{quan2025radiative}, plasma transfer\cite{liu2017unified}, and multiphase systems\cite{LIU2019264}.
Here, UGKS is employed to simulate CRBS using the Strang splitting method to handle the perturbation source term. Compared with DSMC, UGKS allows for larger time steps and cell sizes in near-continuum regimes and achieves higher computational efficiency by eliminating particle noise. On the other hand, simulation efficiency is  enhanced through the Linear Boltzmann Equation (LBE) formulation and Fourier-domain treatment, solved using the fast spectral method\cite{wu2015kinetic} and the general synthetic iteration scheme (GSIS)\cite{su2020fast}. Their results show excellent agreement with both experimental measurements and DSMC simulations, and the approach has been successfully extended to diatomic gases\cite{wu2025coherent}.
While the current method offers significantly higher efficiency than conventional numerical solvers, the key advantage of UGKS lies in its capacity to model larger perturbation accelerations arising from source terms by employing a BGK-type kinetic model instead of a linearized one. Accordingly, this study explores the influence of strong perturbations across different Knudsen numbers (Kn), extending beyond the conventional small-perturbation assumption and addressing physical scenarios similar to those discussed in Refs.\cite{pan_coherent_2003, shneider2007molecular, barker2001optical}.

In this paper, the kinetic equation for CRBS is firstly presented in Sec.\ref{sec: Kinetic equations of crbs}. Then the BGK-Shakhov governing equations and UGKS for CRBS are introduced in Sec.\ref{sec:ugks for crbs}. In Sec.\ref{sec: numerical results}, the numerical method is thoroughly validated by argon experiment, and the influence of incident signal intensity variations on the spectral profile of CRBS under different Kn is investigated. Finally, the conclusion of this study is given in Sec.\ref{sec:conclusion}.


\section{Kinetic equations for CRBS}\label{sec: Kinetic equations of crbs}

Building on the preliminary introduction to the physical mechanism of CRBS in Sec.\ref{Sec: introduction}, this section derives the kinetic equation for CRBS in monatomic gases. Specifically, as illustrated in Fig.\ref{fig: The physical process of CRBS}, two pump beams—both polarized perpendicular to the page—are focused and crossed at their foci to form an interference pattern known as an optical lattice, which induces a wavelike density perturbation field in the gas; a probe beam then undergoes coherent scattering from this perturbation, generating the CRBS signal beam. In general, Einstein’s theoretical framework predicts that the scattering intensity is proportional to the square of dielectric constant fluctuations $(\delta\epsilon)^2$, and for gaseous media, this relationship can be further simplified to show that the scattered light intensity satisfies
\begin{equation}
    I_{\text{sc}} \propto (\delta\rho)^2,
\end{equation}
where $\delta\rho$ denotes density fluctuations\cite{Berne1976Dynamic}. Consequently, the core task of determining the power spectrum of the CRBS signal reduces to solving for the gas density perturbation, which can be obtained by analyzing the kinetic equation:
\begin{equation}
    \frac{\partial f(\boldsymbol{x}, t, \boldsymbol{u})}{\partial t} + \boldsymbol{u} \cdot \nabla f + \boldsymbol{a} \cdot \nabla_{\boldsymbol{u}} f = Q(f, f),
    \label{eq: Boltzman for crbs}
\end{equation}
where $f(\boldsymbol{x}, t, \boldsymbol{u})$ is the microscopic gas distribution function, $\boldsymbol{u}$ is the microscopic velocity, $\boldsymbol{a}$ is the acceleration of a single molecule due to the optical dipole force generated by the crossed pump laser beams. $\boldsymbol{a} \cdot \nabla_{\boldsymbol{u}} f$ is the source term of the perturbation for CRBS, and the collision term $Q(f, f)$ is a quadratic collision operator. As mentioned in Ref.\cite{pan_coherent_2003}, a molecule with polarizability $\alpha$ in an electric field $E$ experiences an optical dipole force given by
\begin{equation}
    F = \frac{1}{2} \alpha \nabla E^2.
\end{equation}
Therefore, the acceleration $a$ of an individual molecule induced by the two single-mode beams is along the $x$-axis and expressed as:
\begin{equation}
    a = \frac{\alpha}{2m} \frac{\partial}{\partial x} \left( {\tilde{\boldsymbol{E}}}_1 + {\tilde{\boldsymbol{E}}}_2 \right)^2 = -\frac{\alpha k E_{10} E_{20}}{2 m} \sin(k x - \omega t),
    \label{eq: acceleration}
\end{equation}
where $m$ is the mass of the gas molecule, $E_{10}$ and $E_{20}$ are the amplitude of the electric field, $k = 2\pi/L$ and $\omega = 2\pi f_d$ are the wave number and angular frequency of the dipole force field, respectively. And $L$ is scattering wavelength, $f_d$ is the frequency difference between the two pump beams.

In the CRBS experimental configuration, the diameter of the interation region is much larger than the dipole force field’s wavelength. Therefore, a plane wave structure of the force field is acceptable. This assumption allows us to use a one-dimensional approximation to describe the amplitudes of the electric fields of two pump lasers\cite{suzuki2024effects}. For the collision term, the kinetic BGK-Shakhov model is employed to characterize it.

The one spatial dimensional BGK-Shakhov model equation can be written as
\begin{equation}
    \frac{\partial f}{\partial t} + u \frac{\partial f}{\partial x} = \frac{f^+ -f}{\tau},
    \label{eq:shakhov bgk eqn}
\end{equation}
where $\tau$ is the relaxation time which is determined by $\tau = \mu /p$ with the macroscopic pressure $p$ and dynamic viscosity  $\mu$. And $f^+$ is defined as follows\cite{shakhov_generalization_1968} to result in a realistic Prandtl number ($\mathrm{Pr}$),
\begin{equation}
\begin{aligned}
    &f^+ = g\left[1 + (1 - \Pr) (u-U) \cdot q \left( \frac{(u-U)^2 + \boldsymbol{\xi^2}}{RT} - 5 \right) / (5pRT) \right],\\
    &g(x, t, u) = \rho (\frac{\lambda}{\pi})^{-\frac{K+1}{2}}  e^{- \lambda [(u-U)^2 + \boldsymbol{\xi} ^2]}.
\end{aligned}
    \label{eq: modified equilibrium distribution}
\end{equation}
In this function, $U$ is the macroscopic velocity, $q$ is heat flux, $R$ is ideal gas constant, and $T$ is the temperature. And $g$ is the local equilibrium Maxwellian distribution, $\lambda = \frac{1}{2 R T}$, $K = 2$ for one diemnsion monatomic gas and $\boldsymbol{\xi}^2 = v^2 + w^2$. When $\mathrm{Pr} = 1$, it goes back to BGK model. The macroscopic variables can be calculated via,
\begin{equation}
\begin{aligned}
    &\boldsymbol{W} =
    \begin{pmatrix}
    \rho \\
    \rho U \\
    \rho E
    \end{pmatrix}
    = \int \boldsymbol{\psi} f \mathrm{d}\boldsymbol{\Xi},  \\
    & q = \frac{1}{2} \int (u-U)[(u-U)^2+ \boldsymbol{\xi}^2]f \mathrm{d}\boldsymbol{\Xi}
\end{aligned}
    \label{eq: conservative}
\end{equation}
where $\rho E$ is total energy, $\boldsymbol{\psi} = (1,\  u,\  1/2(u^2+\boldsymbol{\xi}^2))^{T}$ is the collision invariants and $\mathrm{d} \boldsymbol{\Xi} = \mathrm{d} u \mathrm{d} \boldsymbol{\xi}$ .

To simplify the computation, the small amplitude approximation is adopted for the perturbation source term, which is widely used for numerical simulation\cite{PhysRevLett.89.183001, wu2020accuracy, su2020fast, zeng2023general}, as follows,
\begin{equation}
    \begin{aligned}
        a \frac{\partial f}{\partial u}
        &\approx a  \frac{\partial g_0}{\partial u}    \\
        &= a  \frac{\partial (\rho_0 \left( \frac{\lambda_0}{\pi} \right)^{\frac{K+1}{2}} e^{-\lambda_0 \left( (u - U_0)^2 + \boldsymbol{\xi}^2 \right)})}{\partial u}    \\
         &= -2_0 \lambda_0 (u-U_0) a \rho_0 \left( \frac{\lambda_0}{\pi} \right)^{\frac{K+1}{2}}   ( e^{-\lambda_0 \left( (u - U_0)^2 + \boldsymbol{\xi}^2 \right)}) \\
         &= -2 \lambda_0 (u-U_0) a g_0 , \\
   \end{aligned}
\end{equation}
where $g_0$ is the global equilibrium distribution function. For monatomic gases, suppose the scattering wave vector is propagating along the $x$ direction, the acceleration $a$ is proportional to $\cos(kx - \omega t)$, the CRBS governing equation based on the BGK-Shakhov model, which can be expressed as:
\begin{equation}
    \frac{\partial f}{\partial t} + u \frac{\partial f}{\partial x} = \frac{f^+ - f}{ \tau} + a_0 \cos(kx - \omega t) u g_0 .
    \label{eq:crbs bgk}
\end{equation}

The physics described by Eq.\ref{eq:crbs bgk} corresponds to the dynamics of a gas density perturbation—generated by the source term of the perturbation and relaxed through particle collisions. As the gas transitions from the continuum regime to the rarefied regime, the scattering spectrum evolves: starting from a single-peak Rayleigh scattering, progressing to the coexistence of both Rayleigh and Brillouin components, and finally transforming into a dual-peak Brillouin scattering. Correspondingly, for monatomic gases, the spectral characteristics of the scattered light are characterized by the $\mathrm{Kn}$, which is defined as the ratio of the mean free path $\lambda_{\text{mfp}}$ of gas molecules to the scattering wavelength ($L$):
\begin{equation}
    \text{Kn}=\frac{\lambda_{\text{mfp}}}{L}.
\end{equation}


\section{UGKS for CRBS}\label{sec:ugks for crbs}

\subsection{UGKS for BGK-Shakhov model equation}\label{sec:ugks}
In the general finite volume method (FVM) framework, by using the trapezoidal rule for the approximation of collision term, the BGK-Shakhov model equation (Eq.\ref{eq:shakhov bgk eqn}) becomes,
\begin{equation}
    f_{i,k}^{n+1} = f_{i,k}^n + \frac{1}{\Delta x} \int_{t^n}^{t^{n+1}} u(f_{i-1/2} - f_{i+1/2}) dt + \frac{\Delta t}{2} \left( \frac{f_{i,k}^{+(n+1)} - f_{i,k}^{n+1}}{\tau^{n+1}} + \frac{f_{i,k}^{+(n)} - f_{i,k}^n}{\tau^n} \right).
    \label{eq: finite volume bgk}
\end{equation}
Multiplying the collision invariants to Eq.\ref{eq: finite volume bgk} and integrating over the velocity space, the evolution of conservative variables becomes,
\begin{equation}
    \boldsymbol{W}_i^{n+1} = \boldsymbol{W}_i^n + \frac{1}{\Delta x}(\boldsymbol{F}_{i-1/2} - \boldsymbol{F}_{i+1/2}),
    \label{eq: macroscope equation}
\end{equation}
where
\begin{equation*}
    \boldsymbol{F} = \int_{t^n}^{t^{n+1}} \int \boldsymbol{\psi} u f \mathrm{d} \boldsymbol{\Xi} dt.
\end{equation*}

Take the interface $x_{i+1/2} = 0$ at $t^n = 0$ as an example. An integral solution of the BGK-Shakhov model can be constructed by the method of characteristics\cite{prendergast1993numerical},
\begin{equation}
    f(0, t, u, \boldsymbol{\xi}) = \frac{1}{\tau} \int_0^t f^+(x', t', u, \boldsymbol{\xi}) e^{-(t-t')/\tau} \mathrm{d} t' + e^{-t/\tau} f_0(-u_k t, 0, u, \boldsymbol{\xi}).
    \label{eq: integrate solution}
\end{equation}
In this equation, both collision and free transport are considered about. The first term denotes the accumulation of equilibrium state. The second term denotes the transport of non-equilibrium initial state. Then the integrated flux can be derived as
\begin{equation}
    \int_{t^n}^{t^{n+1}} u f_{i+1/2} \mathrm{d} t = u \left( C_1 f^+_0 + C_2 \frac{\partial f^+}{\partial x} u + C_3 \frac{\partial f^+}{\partial t} + C_4 f_0 + C_5 \frac{\partial f}{\partial x} u \right),
\end{equation}
where,
\begin{equation}
    \begin{aligned}
        C_1 &= \Delta t - \tau \left( 1 - e^{-\Delta t/\tau} \right), \\
        C_2 &= 2\tau^2 \left( 1 - e^{-\Delta t/\tau} \right) - \tau \Delta t - \tau \Delta t e^{-\Delta t/\tau}, \\
        C_3 &= \frac{\Delta t^2}{2} - \tau \Delta t + \tau^2 \left( 1 - e^{-\Delta t/\tau} \right),  \\
        C_4 &= \tau \left( 1 - e^{-\Delta t/\tau} \right), \\
        C_5 &= \tau \Delta t e^{-\Delta t/\tau} - \tau^2 \left( 1 - e^{-\Delta t/\tau} \right) , \\
        \Delta t &= t^{n+1} - t^n . \\
    \end{aligned}
\end{equation}
The reconstruction and the calculation of spatial and temporal derivatives can be referred to Ref.\cite{xu2001gas, xu2010unified, xu2021unified}.

To reduce the computational cost associated with the dimensions $v$ and $w$, the reduced distribution function \cite{yang1995rarefied, chu1965kinetic} is employed as
\begin{equation}
    h = \int f d\boldsymbol{\xi}, \quad b = \int \boldsymbol{\xi}^2 f d\boldsymbol{\xi}.
\end{equation}
Then the governing equation of $f$ using Eq.\ref{eq:shakhov bgk eqn} becomes two similar equations of $h$ and $b$,
\begin{equation}
\begin{aligned}
    \frac{\partial h}{\partial t} + u \frac{\partial h}{\partial x} &= \frac{h^+ -h}{\tau},\\
    \frac{\partial b}{\partial t} + u \frac{\partial b}{\partial x} &= \frac{b^+ -b}{\tau}, \\
\end{aligned}
\end{equation}
where,
\begin{equation}
    \begin{aligned}
        h^+ &= [1+ \frac{4(1 - \text{Pr})\lambda^2}{5\rho}(u - U)q(2\lambda(u - U)^2 + K - 5)] \rho \left( \frac{\lambda}{\pi} \right)^{1/2} e^{-\lambda(u-U)^2}, \\
        b^+ &= \frac{K}{2\lambda} [1+ \frac{4(1 - \text{Pr})\lambda^2}{5\rho}(u - U)q(2\lambda(u - U)^2 + K - 3)] \rho \left( \frac{\lambda}{\pi} \right)^{1/2} e^{-\lambda(u-U)^2} .\\
    \end{aligned}
\end{equation}
As for macroscopic variables, the expressions become
\begin{equation}
    \begin{aligned}
        &\boldsymbol{W} =
        \begin{pmatrix}
        \rho \\
        \rho U \\
        \rho E
        \end{pmatrix}
        = \begin{pmatrix}
        \int h \mathrm{d}u \\
        \int hu \mathrm{d}u \\
        \frac{1}{2}(\int hu^2 \mathrm{d}u + \int b \mathrm{d}u)
        \end{pmatrix}   ,\\
        & q = \frac{1}{2}[\int (u-U)(u-U)^2 h \mathrm{d}u + \int (u-U) b \mathrm{d}u].
    \end{aligned}
\end{equation}


\subsection{Strang splitting method}
Based on Eq.\ref{eq:crbs bgk}, the BGK-Shakhov model for CRBS can be written as
\begin{equation}
    \frac{\partial f}{\partial t} + u \frac{\partial f}{\partial x} = \frac{f^+ - f}{ \tau} + a_0 \cos(2 \pi (x - {f_d}t) u g_0 .
    \label{eq:crbs shakhov}
\end{equation}
Compared with Eq.\ref{eq:shakhov bgk eqn}, this equation incorporates an additional perturbation source term. Therefore, the Strang splitting method\cite{strang1968construction} is considered. This method can effectively decouple the source term from the core solution process of UGKS through symmetric operator decomposition. Moreover, it strictly preserves second-order accuracy for non-commuting operators, enabling compatibility with the second-order UGKS and ensuring the overall numerical scheme retains this level of accuracy.

Based on the Strang splitting method, Eq.\ref{eq:crbs shakhov} during one $\Delta t$ period time could be divided into three individual steps:
\begin{itemize} 
\item Pre-forcing:
\begin{equation} 
\frac{\partial f}{\partial t}  = a_0 \cos[2 \pi (x - {f_d}t)] u g_0,
\label{eq: Acceleration term 1}
\end{equation}
\item UGKS:
\begin{equation}
\frac{\partial f}{\partial t} + u \frac{\partial f}{\partial x}= \frac{f^+ - f}{ \tau},
\label{eq: ugks term}
\end{equation}
\item Post-forcing:
\begin{equation}
\frac{\partial f}{\partial t}  = a_0 \cos [2 \pi (x - {f_d}t)] u g_0
\label{eq: Acceleration term 2}
\end{equation}
\end{itemize}

Owing to the small amplitude approximation, the right-hand side term in the pre-forcing step corresponds to the initial equilibrium Maxwell distribution function. This approximation simplifies the entire process into an inhomogeneous ordinary differential equation (ODE), which can be directly integrated. The calculation formula for the pre-forcing step at the n-th step can be expressed as the following form:
\begin{equation}
    \begin{aligned}
        f^{n^*} - f^n &= \int_{t^n}^{t^{n^*}} a_0 \cos \left[2\pi(x - f_d t)\right] \mathrm{d} t \ u g_0    ,\\
         &= \left. -\frac{a_0}{2\pi f_d} \sin \left[2\pi(x - f_d t)\right] \right|_{t^n }^{t^{n^*}} \ u g_0 , \\
   \end{aligned}
   \label{eq: Acceleration term 1 discretized}
\end{equation}
where
\begin{equation*}
    t^{n^*} =  t^n +\frac{\Delta t}{2}.
\end{equation*}
By taking the moments of the Eq.\ref{eq: Acceleration term 1 discretized}, the macroscopic governing equation can obtain that,
\begin{equation}
\begin{split}
\boldsymbol{W}^{n^*} - \boldsymbol{W}^{n} =  M \
\begin{pmatrix}
0 \\
p_0 \\
0
\end{pmatrix}
\end{split},
\label{eq: Acceleration term 1 macro}
\end{equation}
where
\begin{equation*}
    M = -\frac{a_0}{2\pi f_d}  \ \sin \left[ 2\pi \left( x - f_d t^n - f_d \frac{\Delta t}{2}  \right) \right] + \frac{a_0}{2\pi f_d} \ \sin \left[ 2\pi ( x - f_d t^n ) \right].
\end{equation*}
The computational procedure of the post-forcing step is similar to the pre-forcing one. Then the evolution of the Strang-splitting UGKS for CRBS from time $t^n$ to $t^{n+1}$ is as follows:
\begin{description}
    \item[Step (1)] Pre-forcing. Calculate $f^{n^*}$ and $\boldsymbol{W}^{n^*}$ from $f^n$ and $\boldsymbol{W}^n$ at the cell center according to Eq.\ref{eq: Acceleration term 1 discretized} and Eq.\ref{eq: Acceleration term 1 macro}.
    \item[Step (2)] UGKS. Evolve the $f^{n^*}$ and $\boldsymbol{W}^{n^*}$ to $f^{n^{**}}$ and $\boldsymbol{W}^{n^{**}}$ from time $t^{n}$ to $t^{n+1}$ with a time step $\Delta t$ by implementing the numerical scheme described in Sec.\ref{sec:ugks}.
    \item[Step (3)] Pre-forcing. Update the macroscopic variables $\boldsymbol{W}^{n+1}$ and the distribution function $f^{n+1}$ from $\boldsymbol{W}^{n^{**}}$ and $f^{n^{**}}$ over the time interval spanning $t^{n^*}$ to $t^{n+1}$ by the same treatment as that in the pro-forcing step.
\end{description}

And the time step is determined by the CFL condition,
\begin{equation}
    \Delta t = N_{\text{CFL}} \ \frac{\Delta x}{\text{max} (u_k)},
\end{equation}
where $N_{\text{CFL}}$ is the CFL number, and $u_k$ is the global discrete velocity space.


\subsection{Numerical setup}

In the program, the following nondimensionalization is used,
\begin{equation}
    \begin{aligned}
        &\hat{x} = \frac{x}{L}, \ \hat{\rho} = \frac{\rho}{\rho_0}, \ \hat{T} = \frac{T}{T_0}, \ \hat{p} = \frac{p}{\rho_0 C_0^2}, \\
        &\hat{\mu} = \frac{\mu}{\rho_0 C_0 L_0}, \ \hat{u} = \frac{u}{C_0}, \ \hat{f_d} = \frac{f_d}{C_0/L} , \ \hat{a} = \frac{2a}{C^2_0/L} .
    \end{aligned}
\end{equation}
The free stream variables are related through $C_0 = \sqrt{2 R T_0}$. In the following, all variables are nondimensionalized, but we will drop the `` $\hat{\ }$ " for simplicity, except the section\ref{sec: Pan's s6 Model}.

The spatial domain is from $x = 0$ to wavelength ($L$) of the incident light. And the velocity domain is discretized from $-4C_0$ to $4C_0$ into 40 cells. The initial Maxwellian distribution is calculated by a given molecular number density, temperature, and zero macroscopic velocity. And periodic boundary conditions are employed. In this work, the molecular mean free path is defined by the variable hard-sphere (VHS) model,
\begin{equation}
        \lambda_{\text{mfp}} =  \frac{2(5-2 \omega) (7-2 \omega)}{15} \frac{\mu}{p} \sqrt{\frac{RT}{2 \pi}},
        \label{eq:vss model}
\end{equation}
where the viscosity index $\omega = 0.81$, and the viscosity in the flow field is calculated by $\mu \sim T^{\omega}$.


As the power spectrum of the CRBS can then be calculated from the density perturbation, we record $ \tilde{\rho}(t) = \int_{0}^{1} \delta \rho (x, t) \cos\left( {2\pi x} \right) \mathrm{d}x$ at each time step, and take the Fourier transformation of $\tilde{\rho} (t)$ to obtain the spectrum as Ref.\cite{wu2015kinetic}:
\begin{equation}
    S (f_d) = \left| \int e^{-\mathrm{i} 2\pi f_d t} \tilde{\rho}(t) \mathrm{d}t \right|^2.
    \label{eq: spectrum}
\end{equation}
Since the flow in the time domain exhibits periodic variations, for a given frequency difference $f_d$, we first acquire the time-series data of $\tilde{\rho}(t)$ for each complete period based on the numerical simulation results. The signal corresponding to one complete period is then subjected to a discrete Fourier transform, which converts the time-domain density perturbation component into its frequency-domain representation. Subsequently, the squared $L_2$-norm of the transformed density is computed to obtain the single spectral value for each period. Meanwhile, the convergence criterion error is defined as
\begin{equation}
 \epsilon = \left| \frac {S^{(n+1)}}{S^{(n)}} - 1 \right|,
\end{equation}
where $S^{(n)}$ and $S^{(n+1)}$ denote the $n$-th and $(n+1)$-th computed spectral value for the target $f_d$, respectively. The simulation of each frequency is convergent when the convergence criterion error is less than $1.0 \times 10^{-10}$. When the simulation converges, the corresponding spectral value is recorded. Then $f_d$ is systematically adjusted within the predefined frequency range, and the simulation is rerun to obtain the new spectral values. This iterative scanning process is repeated until the full frequency spectrum is obtained.


\section{Numerical results and discussion} \label{sec: numerical results}


\subsection{Grid independence verification}

The CRBS spectrum exhibits significant differences across varying $\mathrm{Kn}$, making it a typical multiscale problem. For transport-collision decoupling methods such as DSMC method, the physical cell size is limited by the molecular mean free path. Because of the multiscale property of the UGKS, a larger cell size can be employed. Prior to conducting specific computations, grid independence verification is first performed to determine the appropriate grid configuration for the near-continuum regime. On the other hand, this section also demonstrates the benefits of UGKS in terms of cell size reduction.

\begin{figure}[H]
	\centering
	\subfigure[]
        {\label{fig: meshkn0.01}
			\includegraphics[width=0.45 \textwidth]{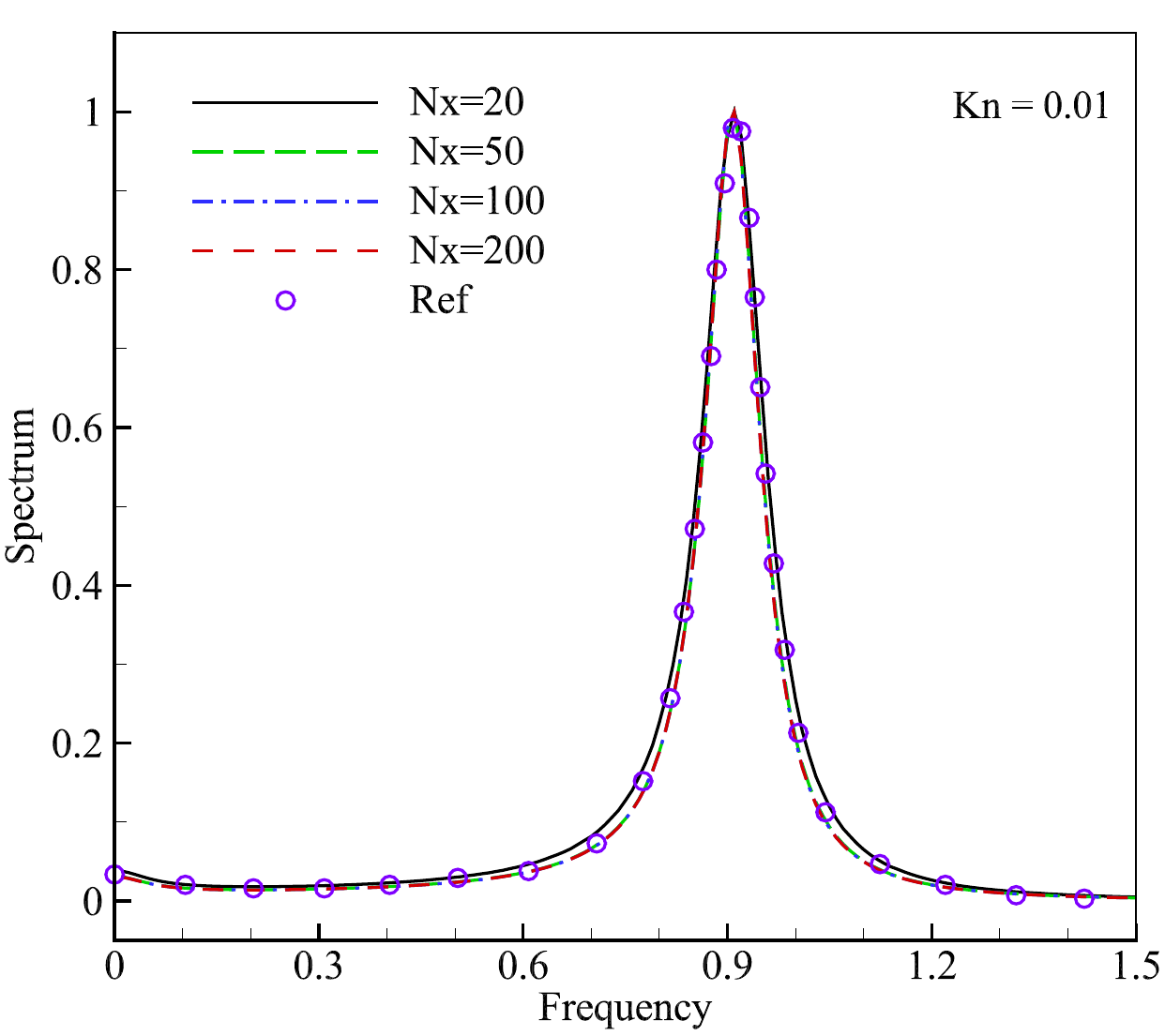}
		}
    \subfigure[]
        {\label{fig: meshkn0.001}
    		\includegraphics[width=0.45 \textwidth]{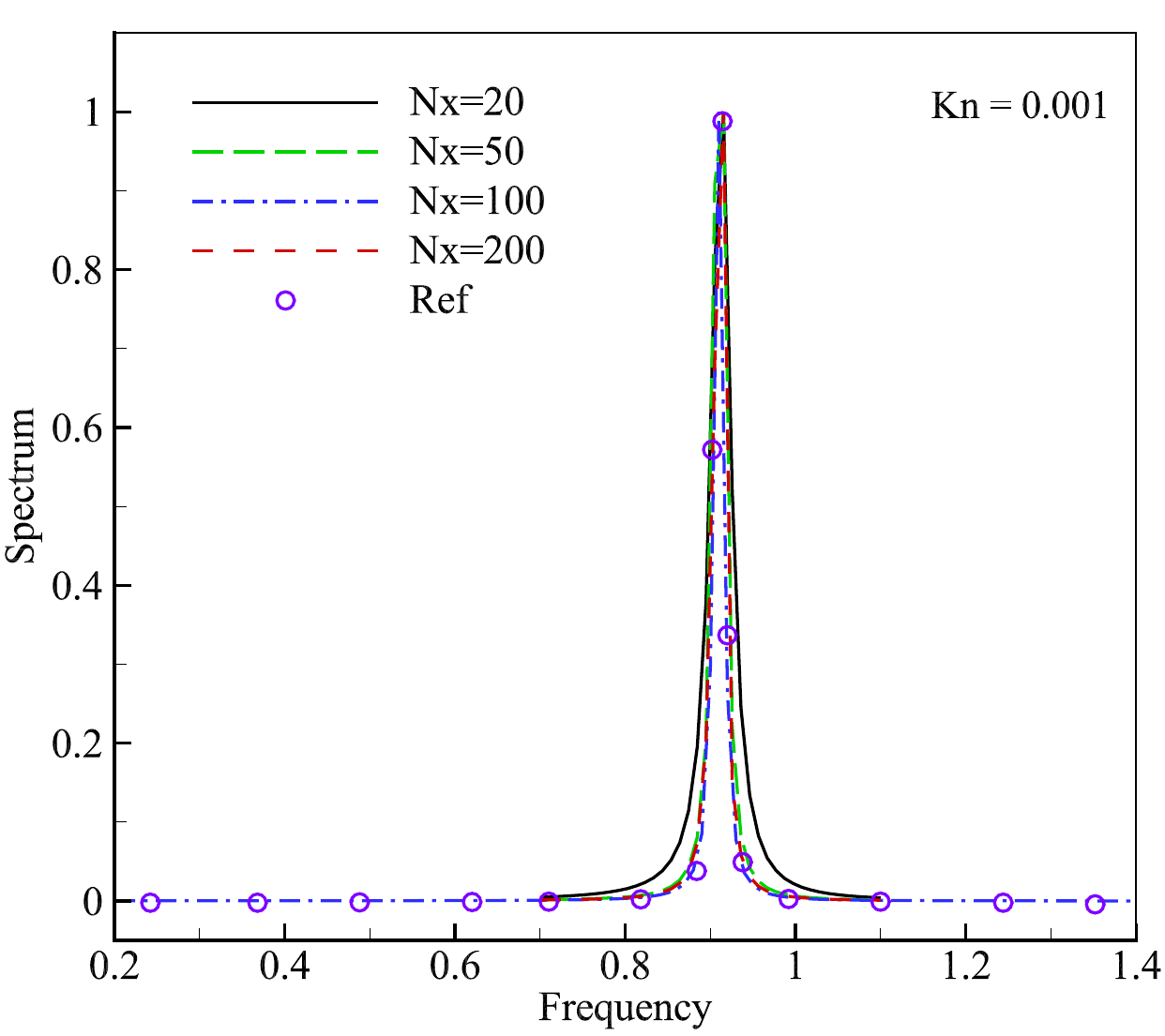}
    	}
	\caption{CRBS spectra of monatomic gas with different cell numbers: (a) $\mathrm{Kn}$ = 0.01; (b) $\mathrm{Kn}$ = 0.001.}
    \label{fig: grid independent test}
\end{figure}

Fig.\ref{fig: grid independent test} presents the CRBS spectrum of argon at $\mathrm{Kn} = 0.01$ and $\mathrm{Kn} = 0.001$, with reference data adopted from Ref.\cite{wu2020accuracy}. The amplitude of acceleration is set to be $a_0 = 0.001$, and UGKS spectra are normalized by the maximum value. As illustrated in this figure, the UGKS results match well with the reference data, and 50 cells are enough for both cases. The physical cell size equals to twice the molecular mean free path at $\mathrm{Kn} = 0.01$, and 20 times molecular mean free path at $\mathrm{Kn} = 0.001$. This observation demonstrates the multiscale capability of the UGKS, which enables reliable simulations across different flow regimes without strict grid constraints.

\subsection{Validation case on Argon experiment} \label{sec: Pan's s6 Model}

The UGKS results on line shape in argon are compared with experimental data and analytical solutions in this section. The experimental measurement is obtained using nonchirped CRBS, and the analytical solutions are obtained from Pan’s s6 (six moment) model\cite{PhysRevLett.89.183001, pan_coherent_2003}. The Pan's s6 model solutions are obtained using the linearized BGK kinetic equation, which is transformed into temporal and spatial frequency domain using Fourier transform. The wavelength of the two pump lasers is $L = 532 nm$, and $k= 2.36 \times 10^7\ rad/m$. The CRBS line shape spans a  range of about 6 GHz. Thus the frequency difference $f_d$ is linearly varied from -3 GHz to 3 GHz. The amplitude of the acceleration varies less than $2\%$ over the interested range except for the pump laser’s longitudinal mode structure, so it's a constant. And for argon, $a_0 = 7.44 \times 10^9\ m/s^2$. The shear viscosity data is obtained by linear interpolation at corresponding temperature $T_0 = 292 \mathrm{K}$. For argon, $\mu = 22.39 \times 10^{-6} \ \text{Pa} \cdot s$.

In Pan's model for atomic gases, the $y$ parameter is defined to characterize kinetic regime using the shear viscosity $\mu$ by\cite{tenti1974kinetic}

\begin{equation}
    y = \frac{1}{k C_0\tau} = \frac{8}{3\sqrt{2}\pi} \frac{\rho \sqrt{R T}}{\mu k},
\end{equation}
where $\tau$ is the collision time which is constant for these cases. As can be inferred from the definition, $y$ exhibits an inverse proportionality to the $\mathrm{Kn}$: the more rarefied the gas, the smaller the value of $y$; conversely, the more continuous, the larger $y$ becomes. Furthermore, in the experiments, the rarefaction degree of the flow field is primarily adjusted by changing the pressure. Specifically, an increase in pressure results in a corresponding rise in the $y$.

\begin{figure}[H]
	\centering
	\subfigure[]
        {\label{fig: y0.023}
			\includegraphics[width=0.3 \textwidth]{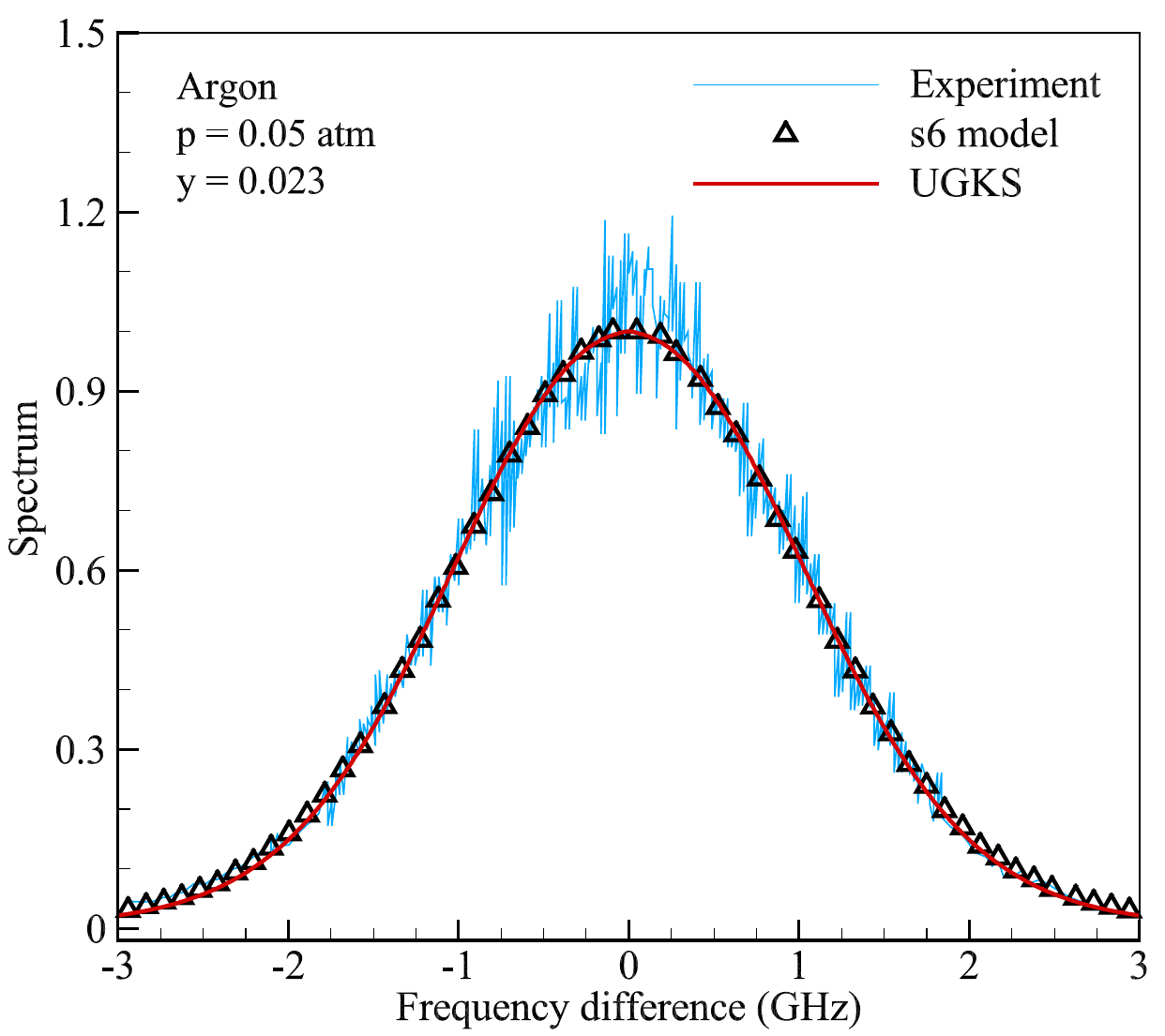}
		}
    \subfigure[]
        {\label{fig: y0.47}
    		\includegraphics[width=0.3 \textwidth]{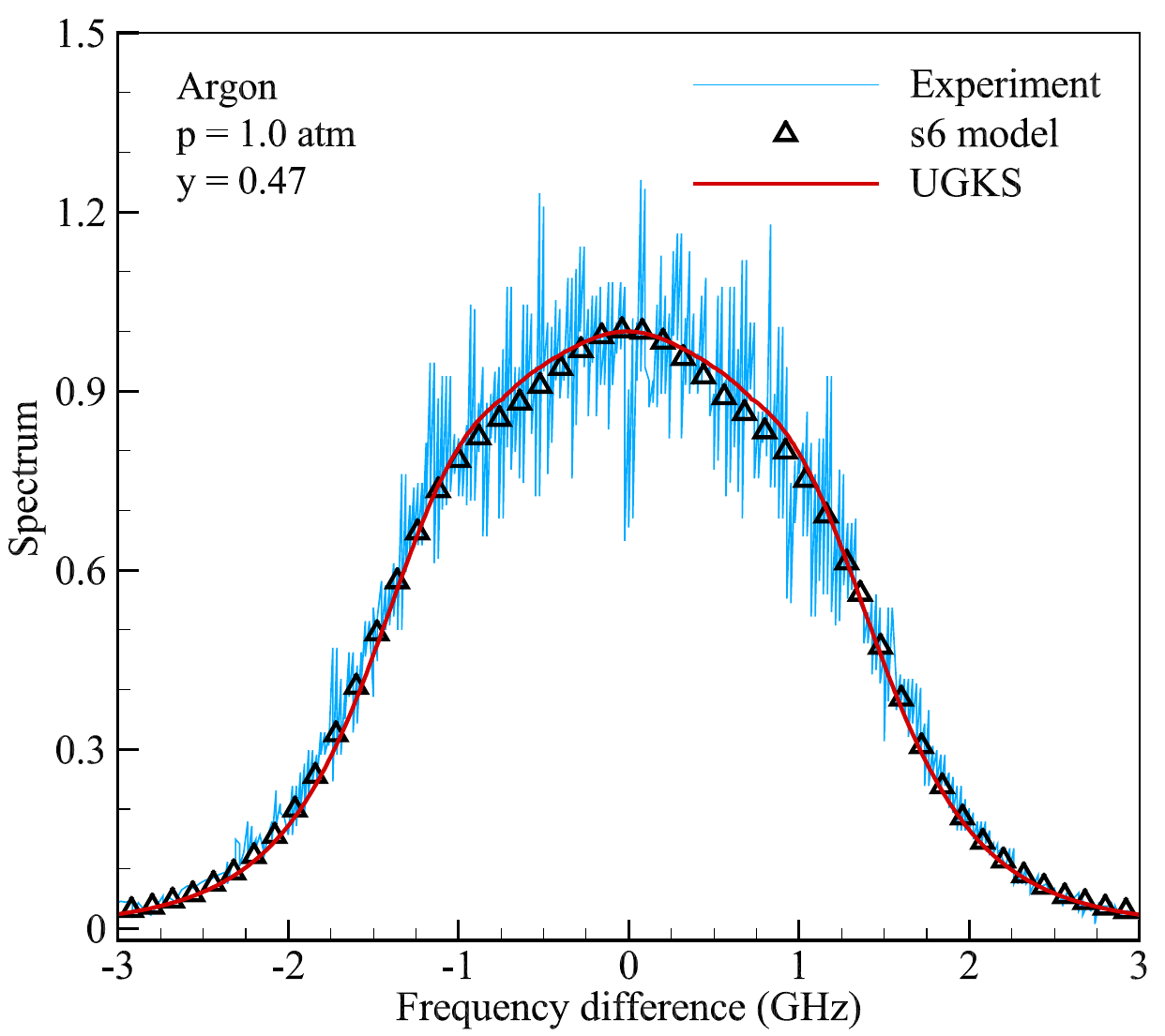}
    	}
    \subfigure[]
        {\label{fig: y0.93}
    		\includegraphics[width=0.3 \textwidth]{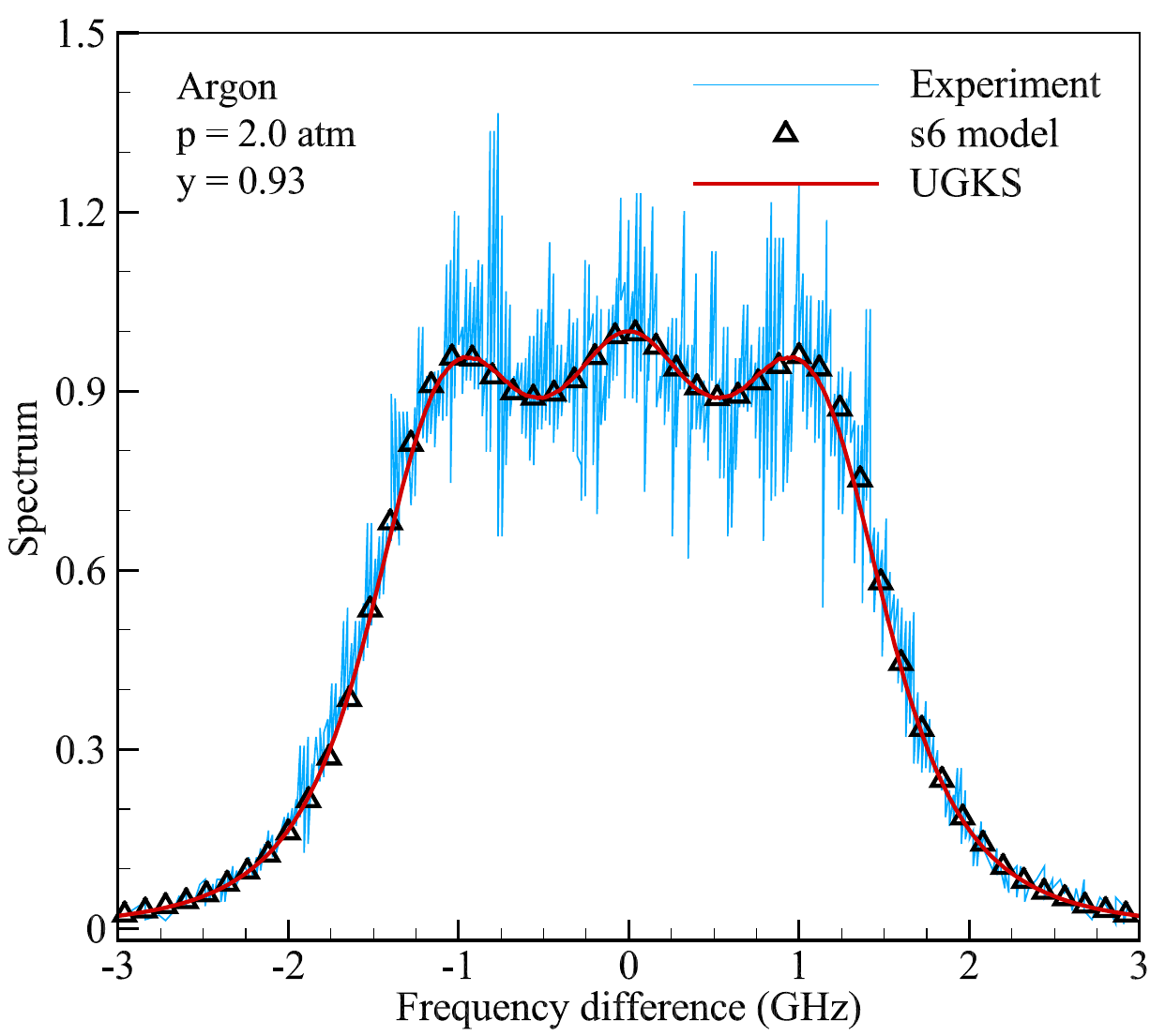}
    	}

    \subfigure[]
        {\label{fig: y1.4}
    		\includegraphics[width=0.3 \textwidth]{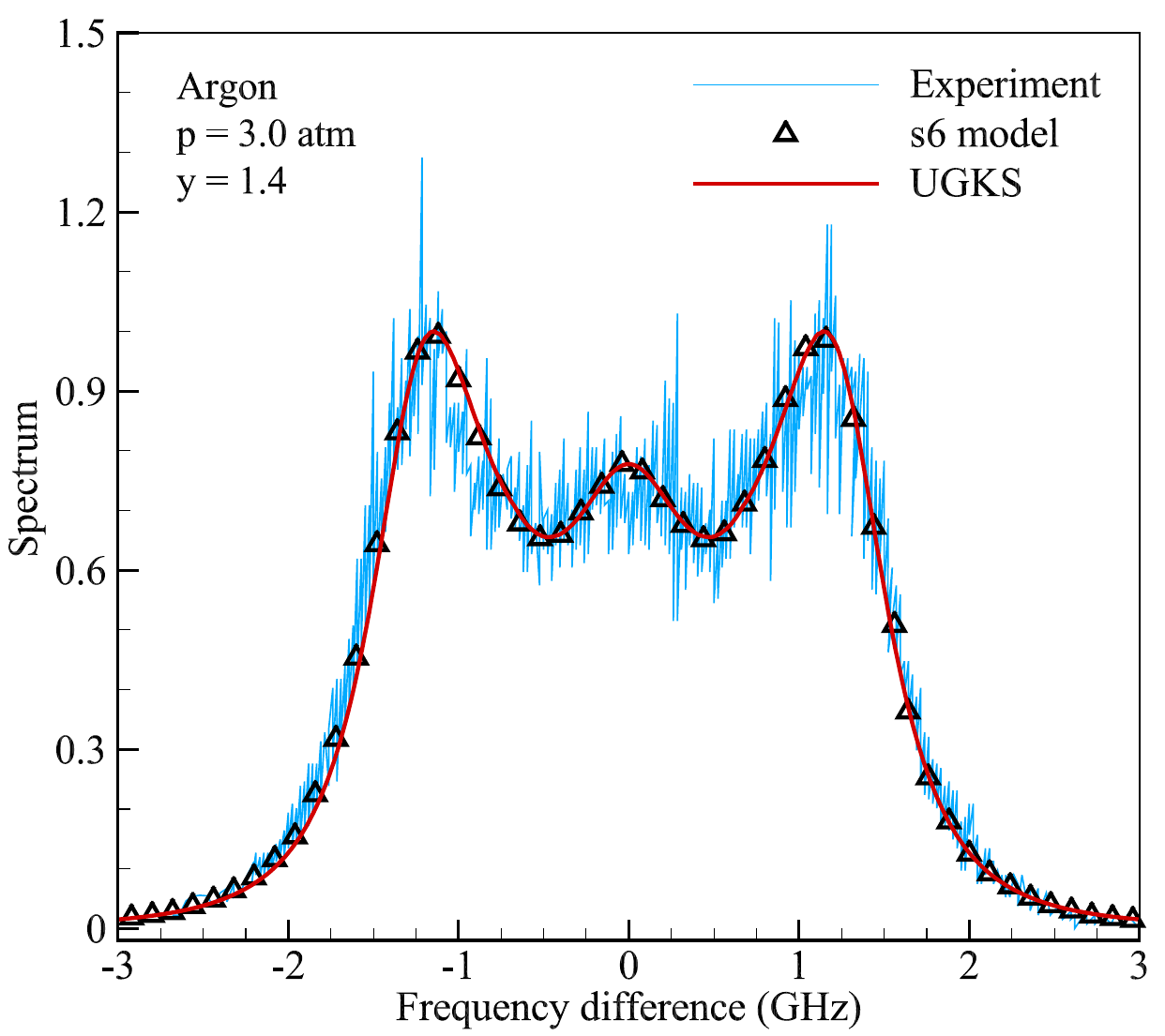}
    	}
    \subfigure[]
        {\label{fig: y1.9}
    		\includegraphics[width=0.3 \textwidth]{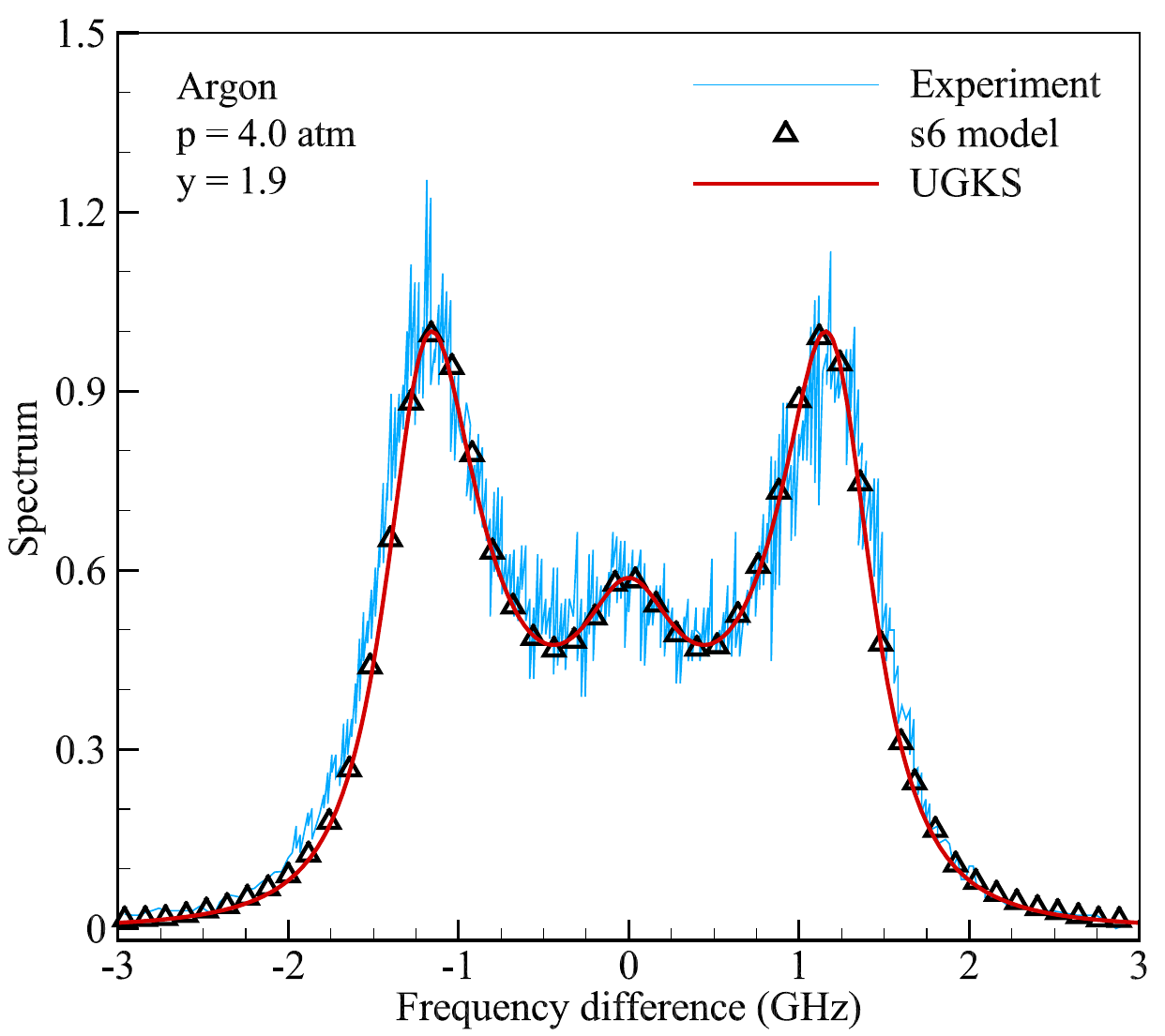}
    	}
    \subfigure[]
        {\label{fig: y2.3}
    		\includegraphics[width=0.3 \textwidth]{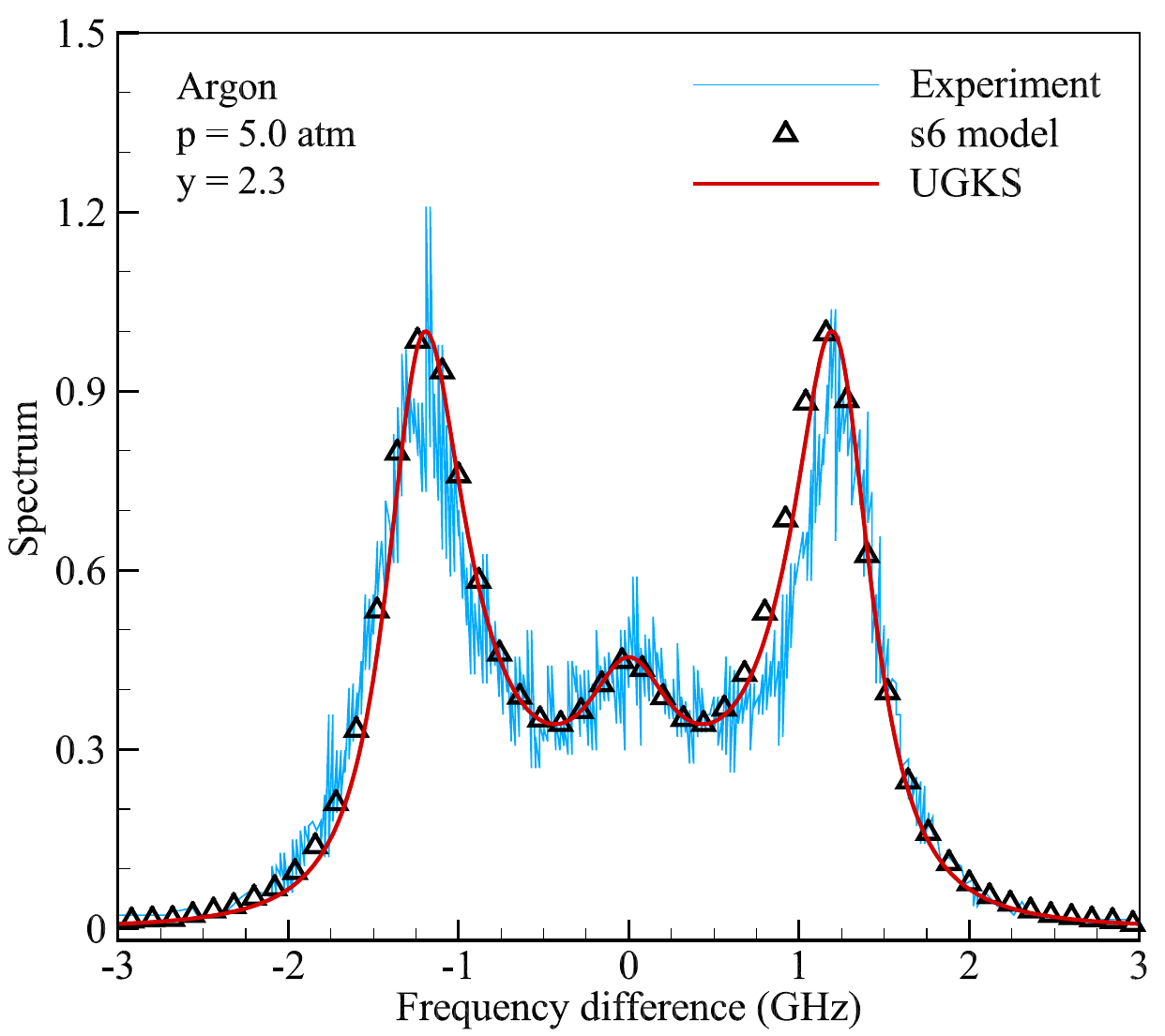}
    	}
	\caption{Power spectrum of CRBS at $T_0 = 292 \mathrm{K}$ in argon.}
    \label{fig: argon experiment}
\end{figure}

Fig.\ref{fig: argon experiment} shows the CRBS line of the argon at different parameters $y$. The UGKS spectra are normalized by the maximum value. As shown in the figures, the UGKS results agree well with both the experiment and analytical solutions across different $y$. When $y$ is small, as shown in the Fig.\ref{fig: y0.023} and \ref{fig: y0.47}, only a single Rayleigh peak is shown, centered around 0 GHz. This line shape indicates that it is in the coherent Rayleigh scattering (CRS) regime where the Rayleigh peak is dominant. However, it can also be seen that with $y$ increasing there is a finite effect of the two Brillouin peaks at the sides of the Rayleigh peak indicated by red solid line when $y = 0.47$. The Brillouin peaks correspond to $\omega = k v_s$, where $v_s = \sqrt{\gamma R T_0}$ is the speed of sound and $\gamma$ is the heat capacity ratio. Fig.\ref{fig: y0.93} shows that the two Brillouin peaks become comparable to the Rayleigh peak when $y = 0.93$. And as $y$ increases, the amplitude of the Brillouin peaks becomes larger than that of the Rayleigh peak. The higher pressure case in Fig.\ref{fig: y2.3} shows larger Brillouin peaks, which indicates a transition to coherent Brillouin scattering (CBS)\cite{suzuki2024effects}.

\subsection{CRBS spectra with different \rm{Kn}}

\begin{figure}[H]
    \centering
    \includegraphics[width=0.5\linewidth]{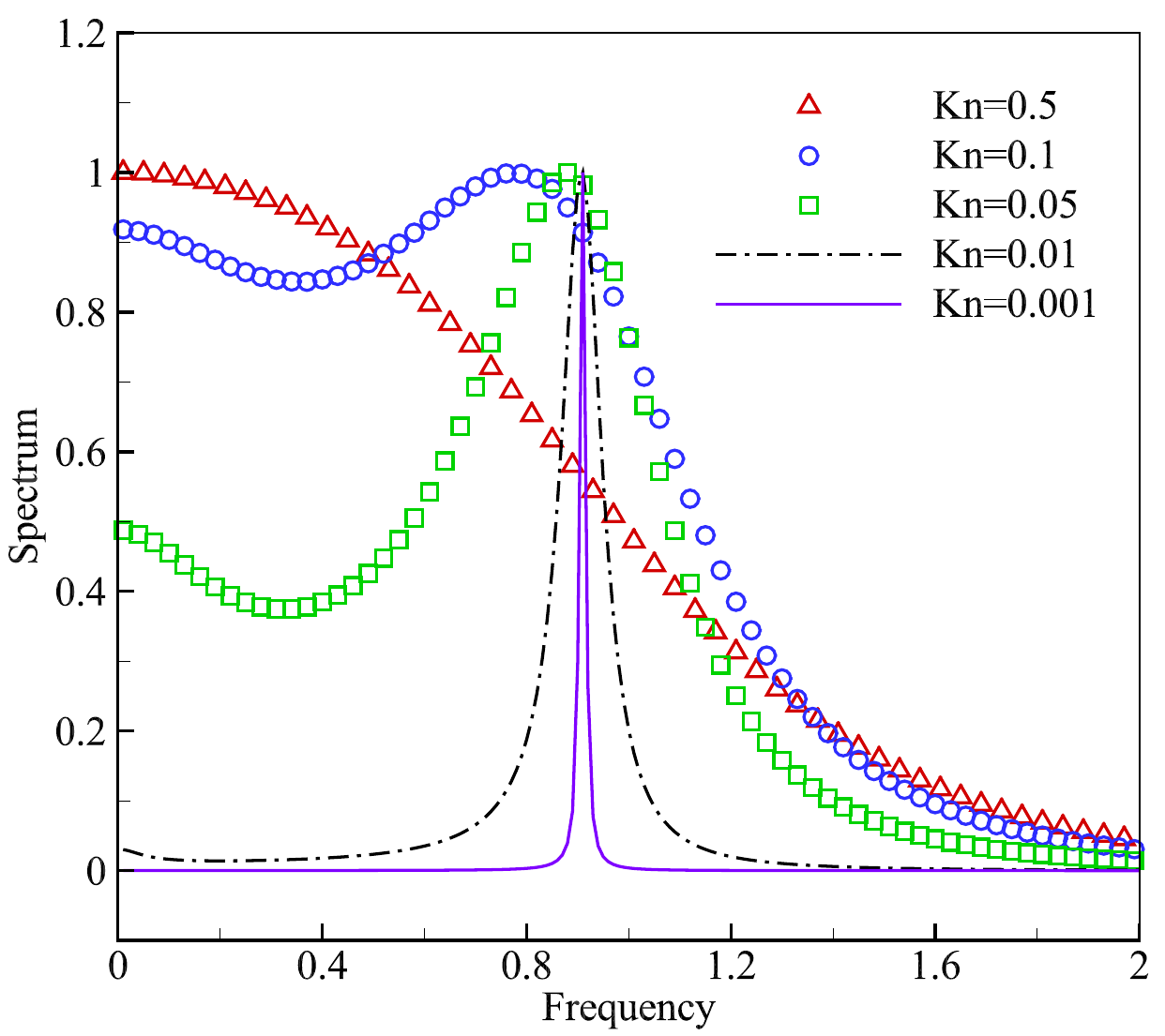}
    \caption{Spectra of the CRBS from the UGKS for monatomic gases with different $\mathrm{Kn}$.}
    \label{fig: kn_change}
\end{figure}

After the validation by experiment data and s6 model, in this section, we  study the CRBS in a further larger Kn range. Fig.\ref{fig: kn_change} depicts typical spectra of the CRBS when the frequencies difference ($f_d$) are larger than zero, and the left part of the spectrum is symmetric with the right one. When $\mathrm{Kn} = 0.001$, the spectrum consists of the Brillouin peaks only. The Brillouin peaks locate at $f_d = \pm v_s / C_0 = \pm \sqrt{\gamma / 2}$. Since $v_s$ is the  sound speed, this spectrum gives a strong evidence that its Brillouin part is related to the  acoustic effect of gas molecules. When $\mathrm{Kn} = 0.01$, there is an important difference that the coherent power spectrum has stronger Brillouin peaks and a much weaker Rayleigh peak located at $f_d = 0$, indicating that the isentropic sound waves are resonantly pumped by the optical dipole force field. As $\mathrm{Kn}$ increases, the relative intensity of the Rayleigh peak to the Brillouin side peaks increases. The spectra consist of the central Rayleigh part and two Brillouin side peaks. Furthermore, the local extrema around the Brillouin peak obtained from UGKS are found at a frequency slightly below the $f_d = \sqrt{\gamma / 2}$ due to the superposition of the comparable peaks.

Eventually the whole spectrum is Gaussian, which  means that the spectrum is solely due to the random thermal motion of gas molecules, when $\mathrm{Kn} = 0.5$. Note that normally the gas is in the free molecular flow regime where the binary collision is negligible when $\mathrm{Kn} = 10$\cite{wu2020accuracy}, here again we observe the Gaussian spectrum (a signature of free molecular flow) when $\mathrm{Kn} = 0.5$, and this spectrum will keep the similar profile with $\mathrm{Kn}$ increasing as shown in the Fig.\ref{fig: y0.023}.

\subsection{Effect of laser intensity on CRBS spectra}

\begin{figure}[H]
	\centering
	\subfigure[]
        {\label{fig: a_kn0.05}
			\includegraphics[width=0.3 \textwidth]{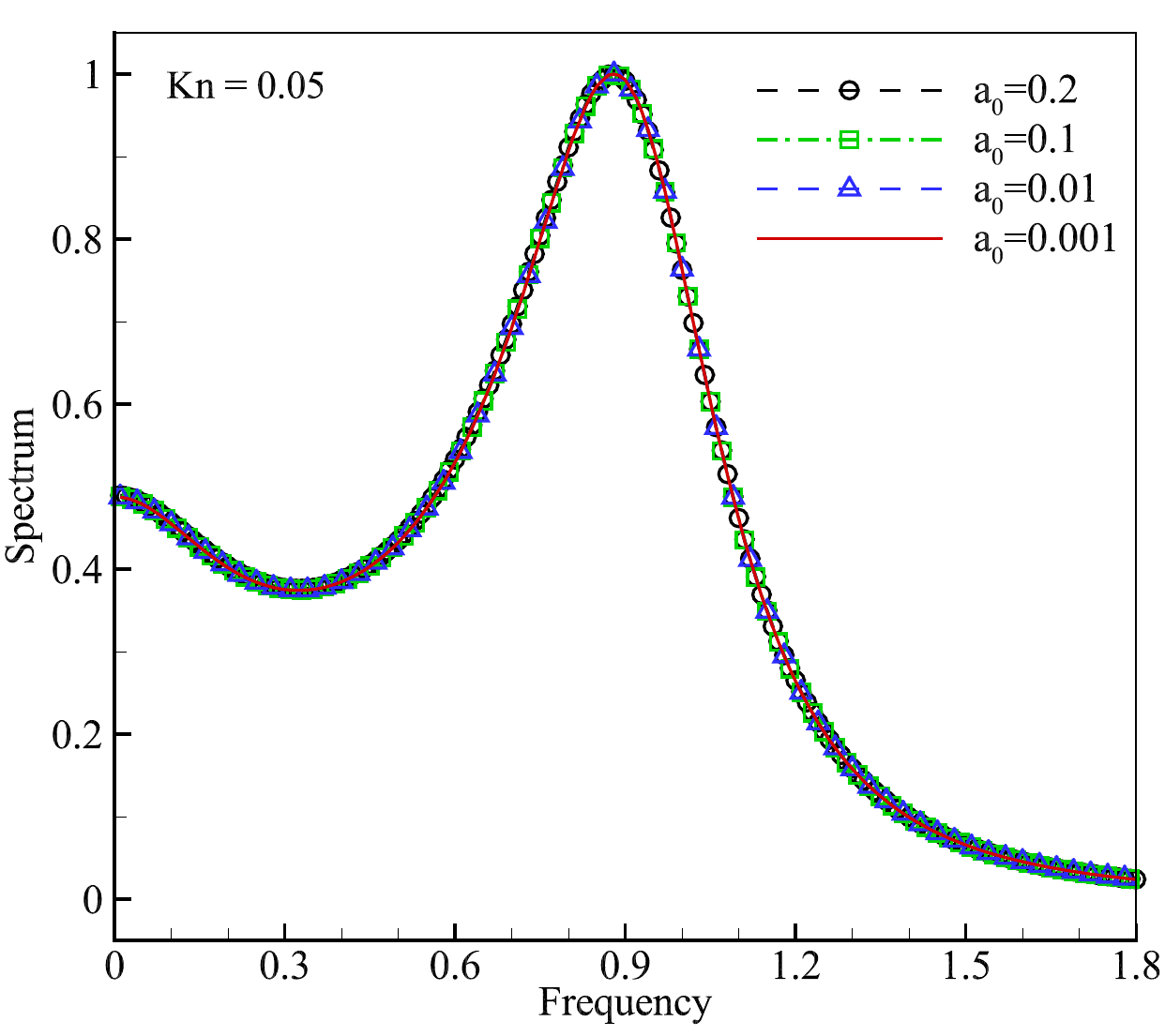}
		}
    \subfigure[]
        {\label{fig: a_kn0.01}
    		\includegraphics[width=0.3 \textwidth]{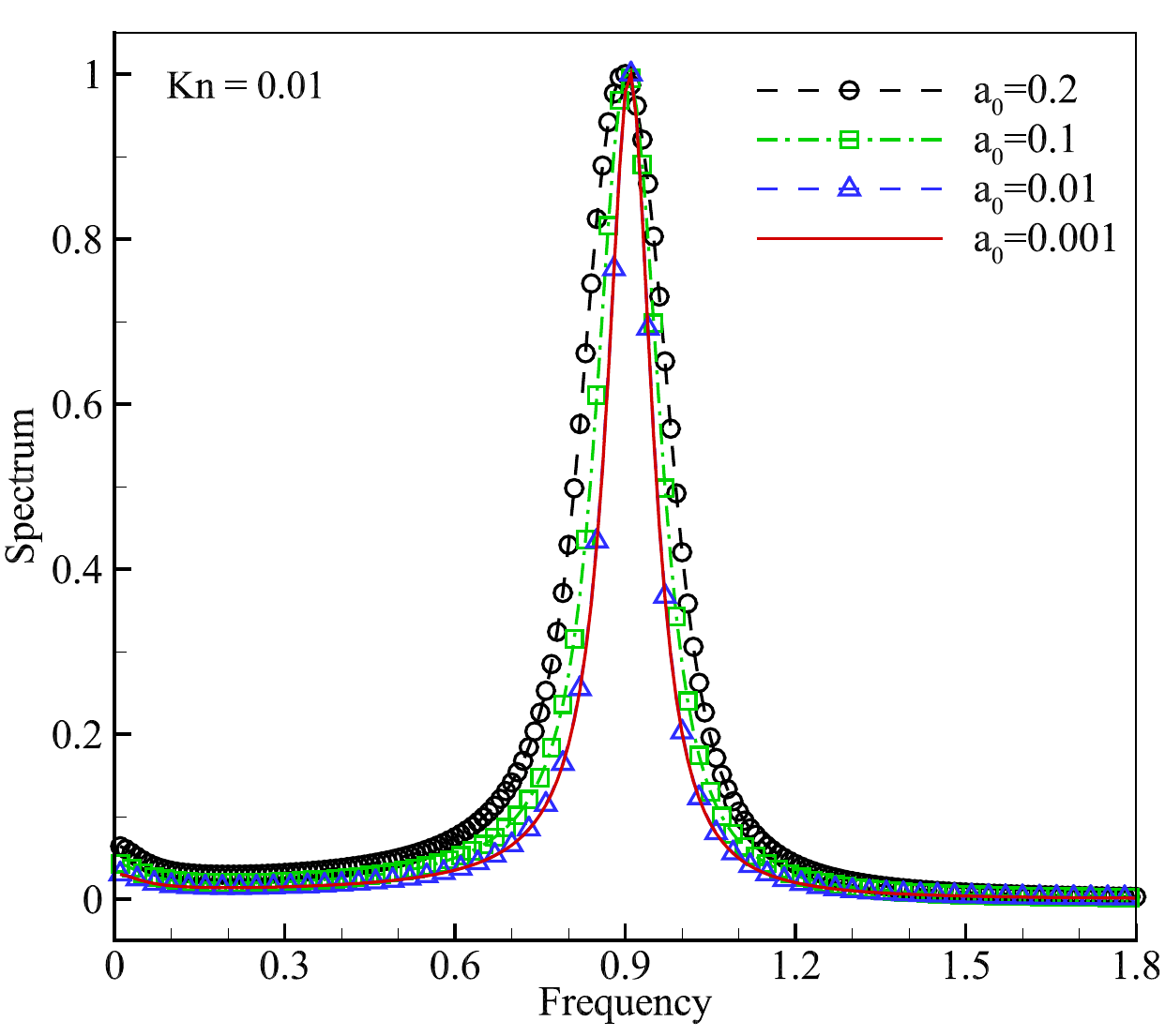}
    	}
    \subfigure[]
        {\label{fig: a_kn0.001}
    		\includegraphics[width=0.3 \textwidth]{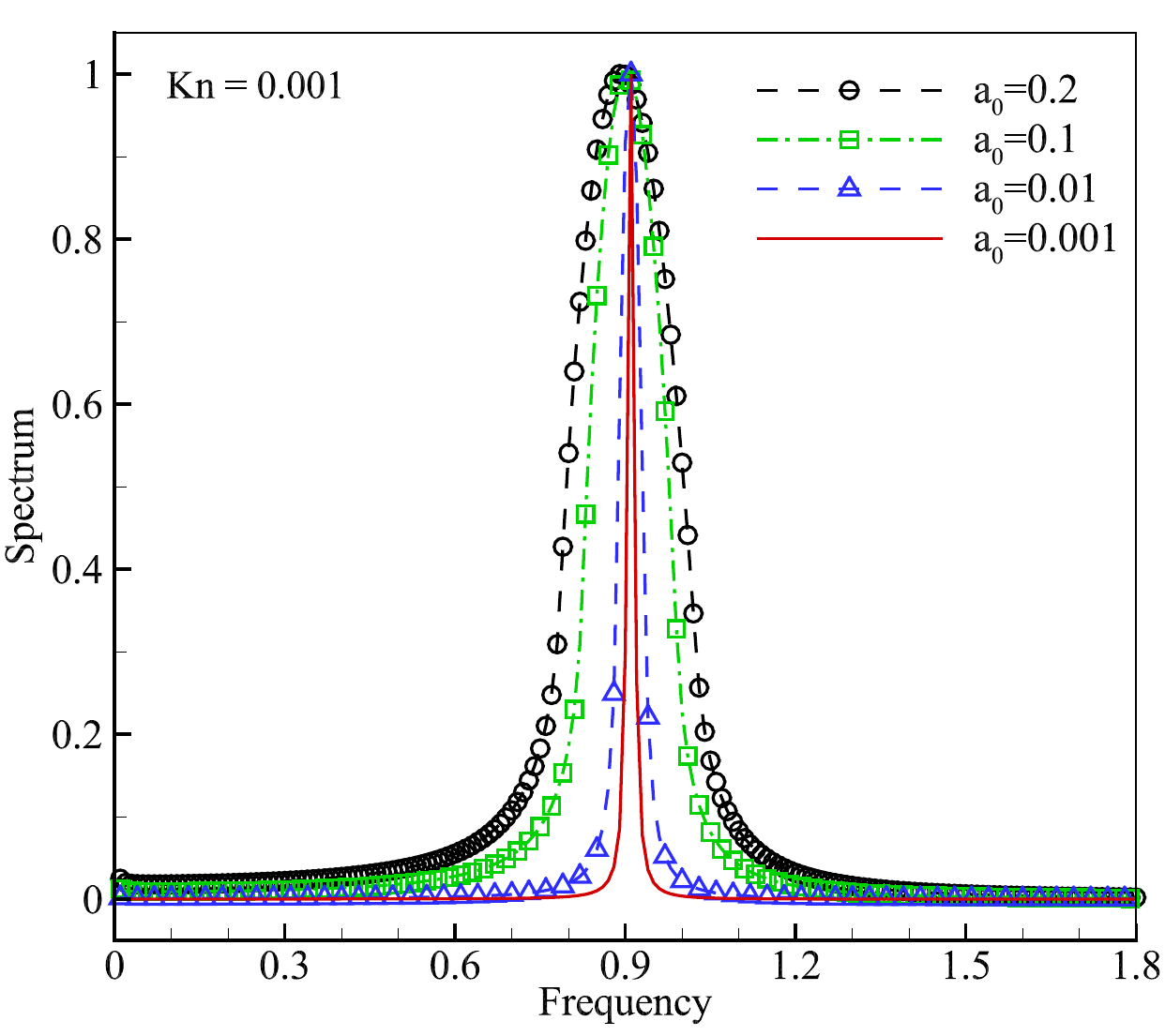}
    	}
	\caption{CRBS spectra with different acceleration amplitude.}
    \label{fig: a0 spectrum}
\end{figure}

As can be seen from Eq.\ref{eq: acceleration}, the acceleration amplitude $a_0$ in the perturbation source term depends on the intensity of the incident laser. Accordingly, this section investigates the influence of variations in $a_0$ on the spectra for different $\mathrm{Kn}$. The CRBS spectra under different accelerations corresponding to $\mathrm{Kn}$ 0.05, 0.01, and 0.001 are presented in Fig.\ref{fig: a0 spectrum}. Overall, compared to the original spectrum, the increase in the acceleration amplitude ($a_0$) leads to a slight offset to the calculated CRBS spectrum, and the resulting line shape resembles that of higher Knudsen number conditions, which indicates the transition from Brillouin scattering to Rayleigh scattering. The comparison of Fig.\ref{fig: a_kn0.01} and Fig.\ref{fig: a_kn0.001} reveals that in the case of only Brillouin peaks, varying the acceleration magnitude yields spectral profiles with similar shapes, with only the peak width changing.

Meanwhile, when the acceleration amplitude decreases to a certain threshold, the CRBS spectrum ceases to change further. As $\mathrm{Kn}$ number decreases, the value of the threshold decreases. For example, when $\mathrm{Kn} = 0.05$, the calculation results obtained with $a_0 > 0.2$ are identical; in contrast, at $\mathrm{Kn} = 0.001$, the spectrum remains unchanged only after $a_0$ is reduced to 0.001. The underlying reason for this phenomenon lies in the fact that fluids in the continuum flow regime exhibit low viscosity, which results in weak resistance to perturbations. Consequently, even a small acceleration tends to alter the flow structure more easily. It can be concluded from the figure that in the continuum flow regime, the smaller the length of the molecular mean free path, the smaller the acceleration amplitude required to maintain a constant spectrum. However, this does not imply that a smaller $a_0$ is always more favorable. In Section\ref{sec: Pan's s6 Model}, the non-dimensionalized $a_0$ corresponding to the argon experiment is 0.065, which is much larger than the minimum $a_0$ of 0.001 chosen in the numerical cases herein. This indicates that if the operating condition of the experimental measurement is set to $\mathrm{Kn} = 0.01$ or $\mathrm{Kn} = 0.001$, the CRBS spectrum of argon may lie between the green line and the blue line in Fig.\ref{fig: a_kn0.01} and Fig.\ref{fig: a_kn0.001} respectively.

To further explore the mechanism by which acceleration amplitude affects the spectral profile, flow field results corresponding to different acceleration amplitudes ($a_0$) at $\mathrm{Kn} = 0.05$, 0.01, and 0.001 are selected for analysis. Since the frequency of $\sqrt{5/6}$ corresponds to the Brillouin peaks of monatomic gases, selecting this frequency ensures more pronounced results in the near-continuum flow regime.

\begin{figure}[H]
	\centering
	\subfigure[]
        {\label{fig: rho_x_a0.01}
    		\includegraphics[width=0.3 \textwidth]{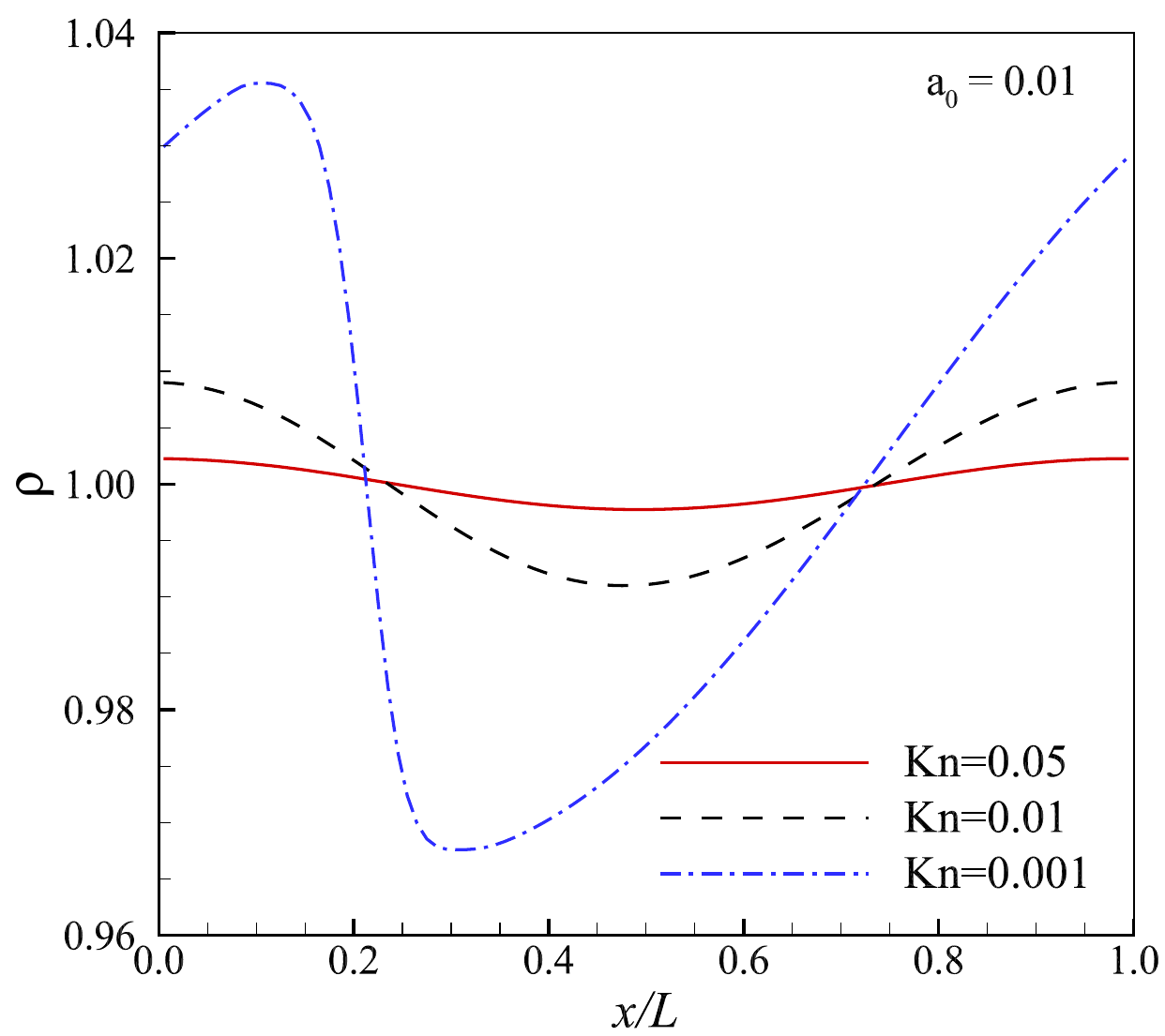}
    	}
    \subfigure[]
        {\label{fig: rho_x_a0.1}
    		\includegraphics[width=0.3 \textwidth]{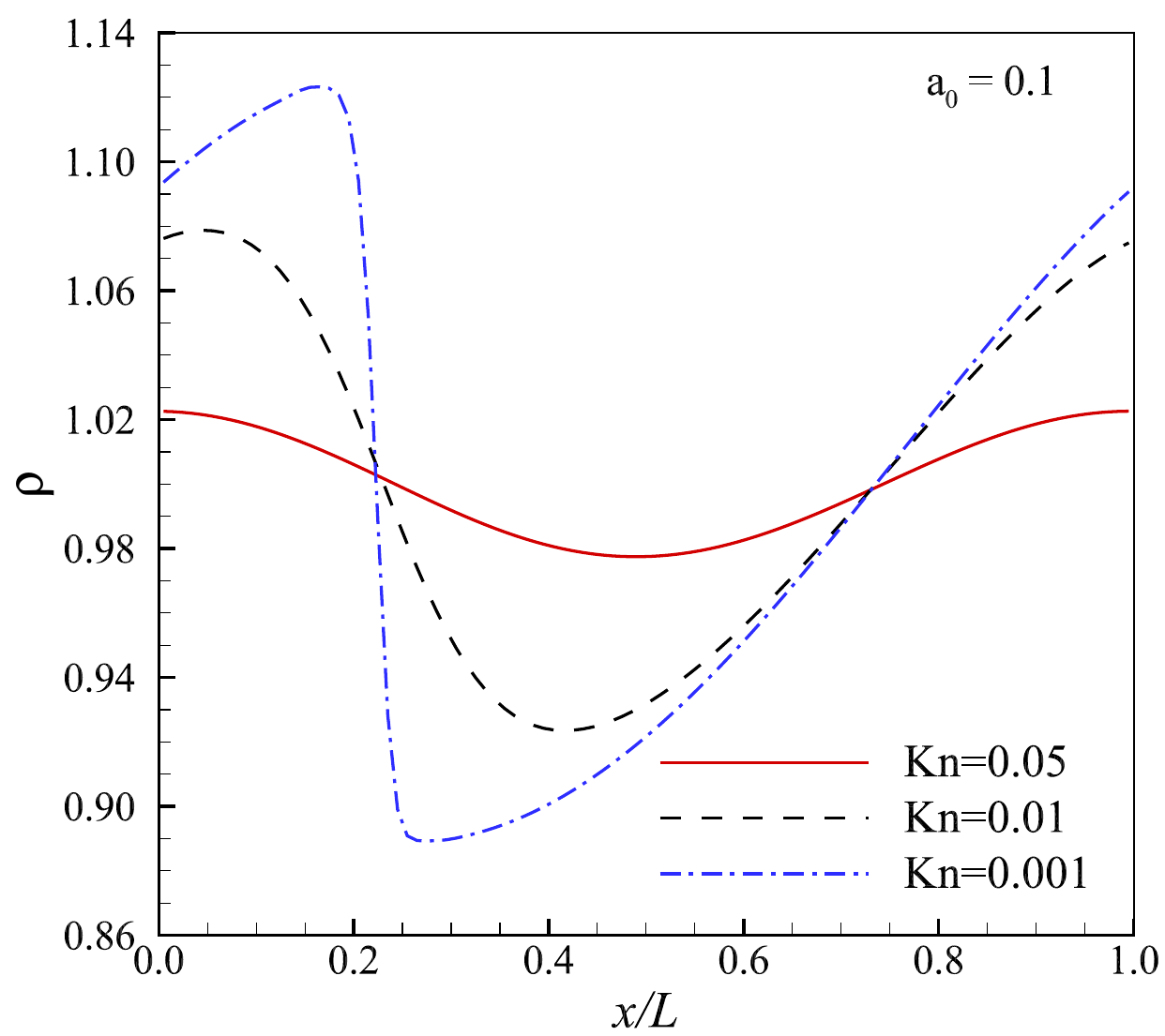}
    	}
    \subfigure[]
        {\label{fig: rho_x_a0.2}
			\includegraphics[width=0.3 \textwidth]{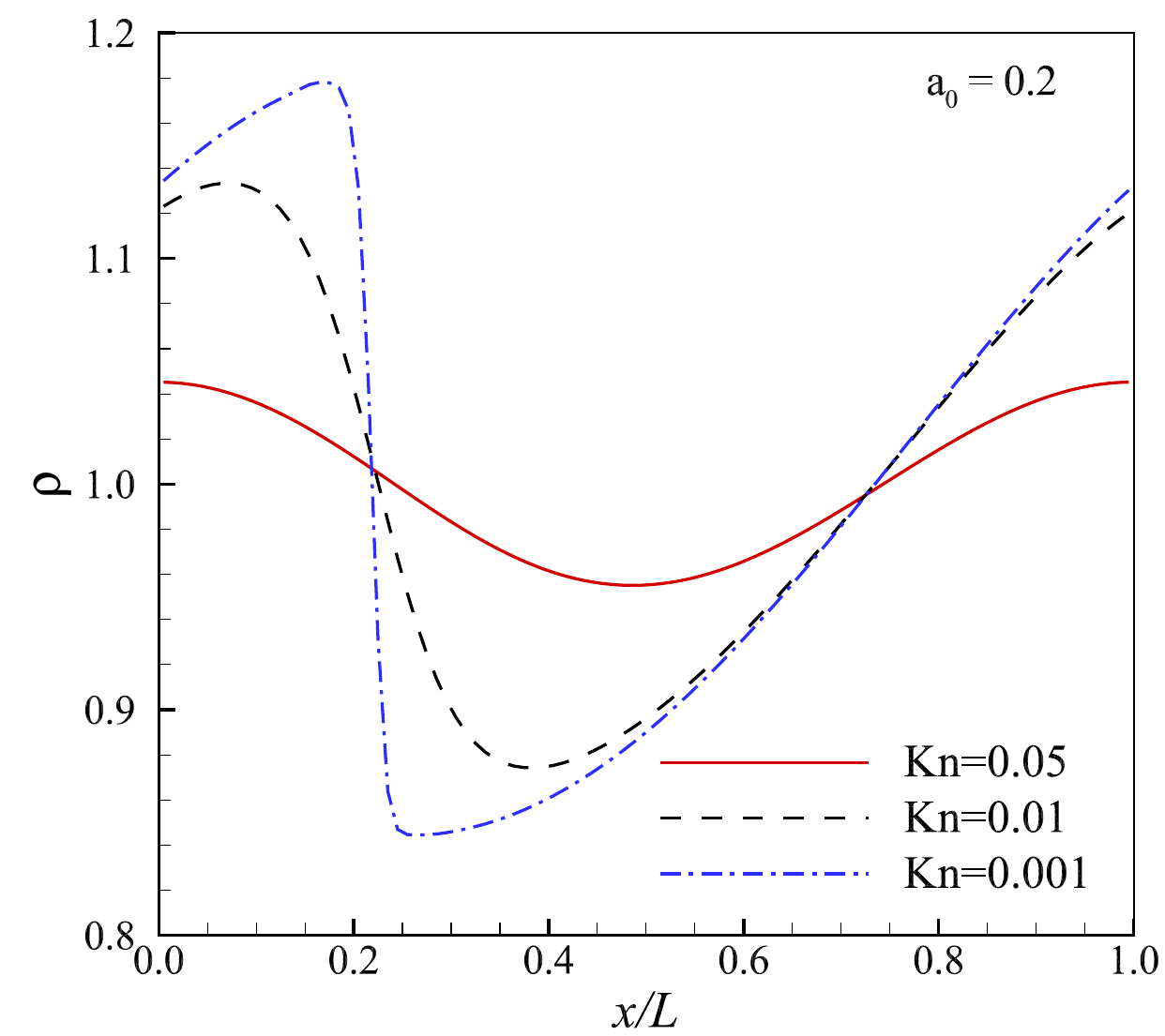}
		}

    \subfigure[]
        {\label{fig: u_x_a0.01}
    		\includegraphics[width=0.3 \textwidth]{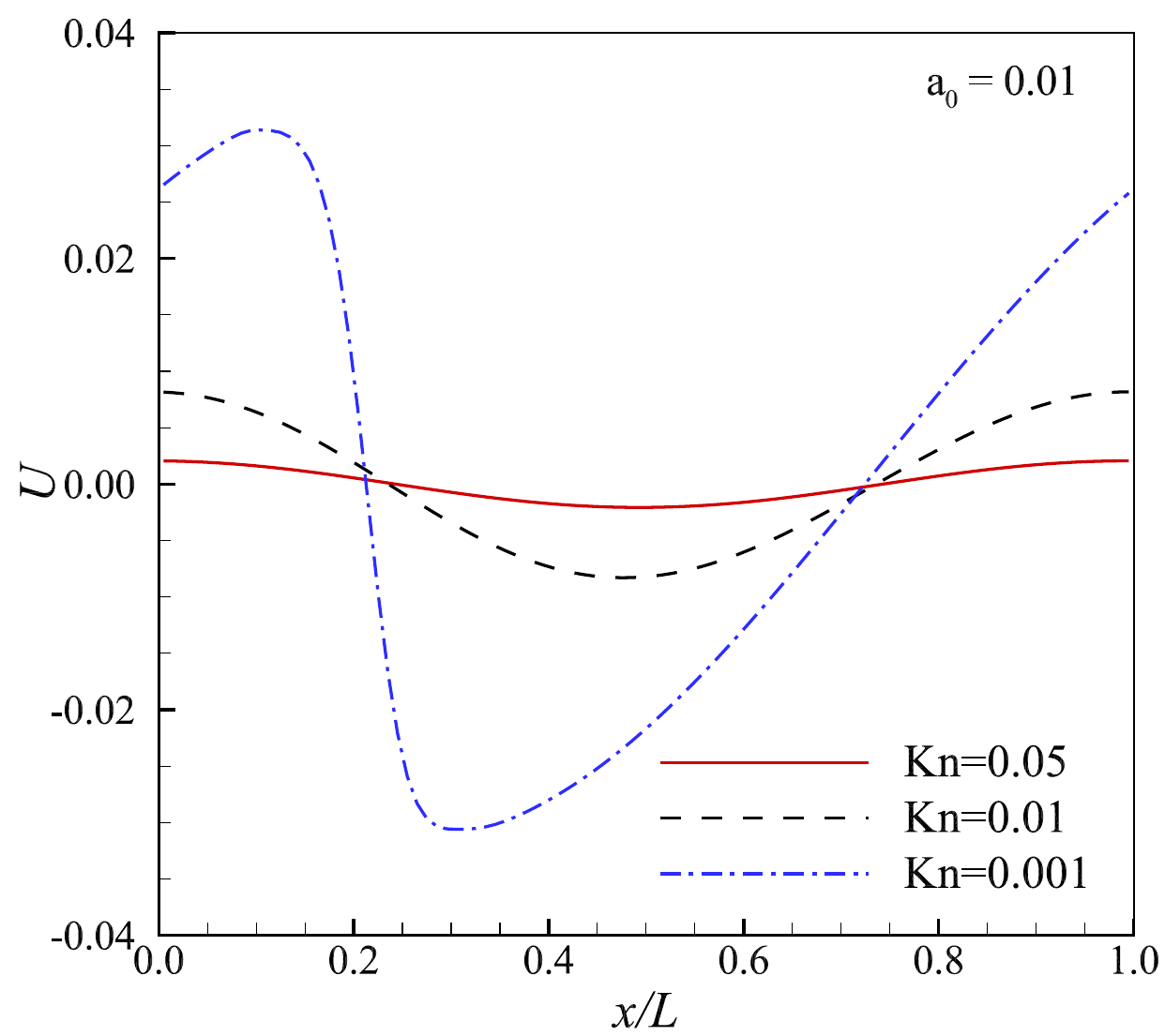}
    	}
    \subfigure[]
        {\label{fig: u_x_a0.1}
    		\includegraphics[width=0.3 \textwidth]{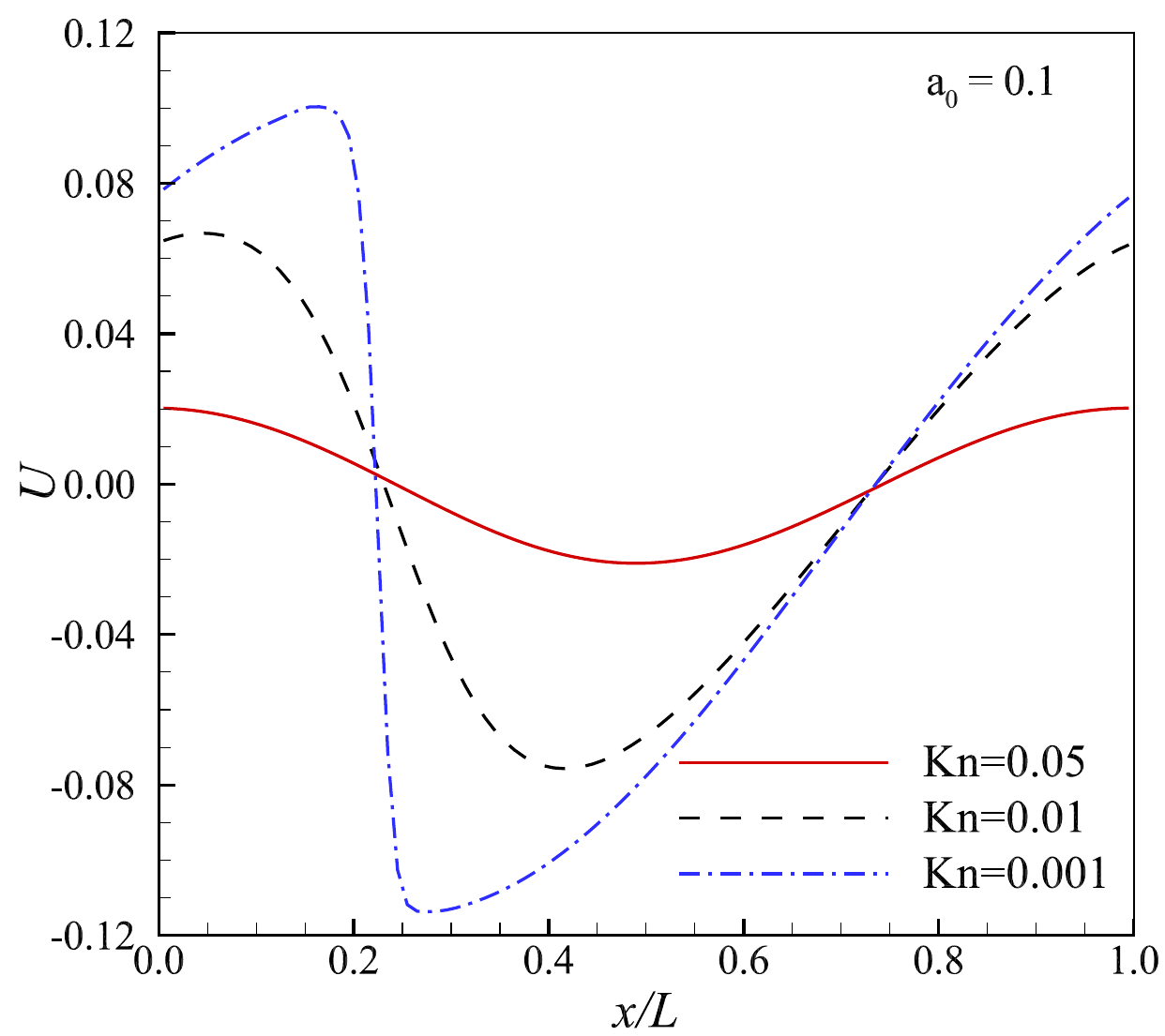}
    	}
    \subfigure[]
        {\label{fig: u_x_a1.0}
    		\includegraphics[width=0.3 \textwidth]{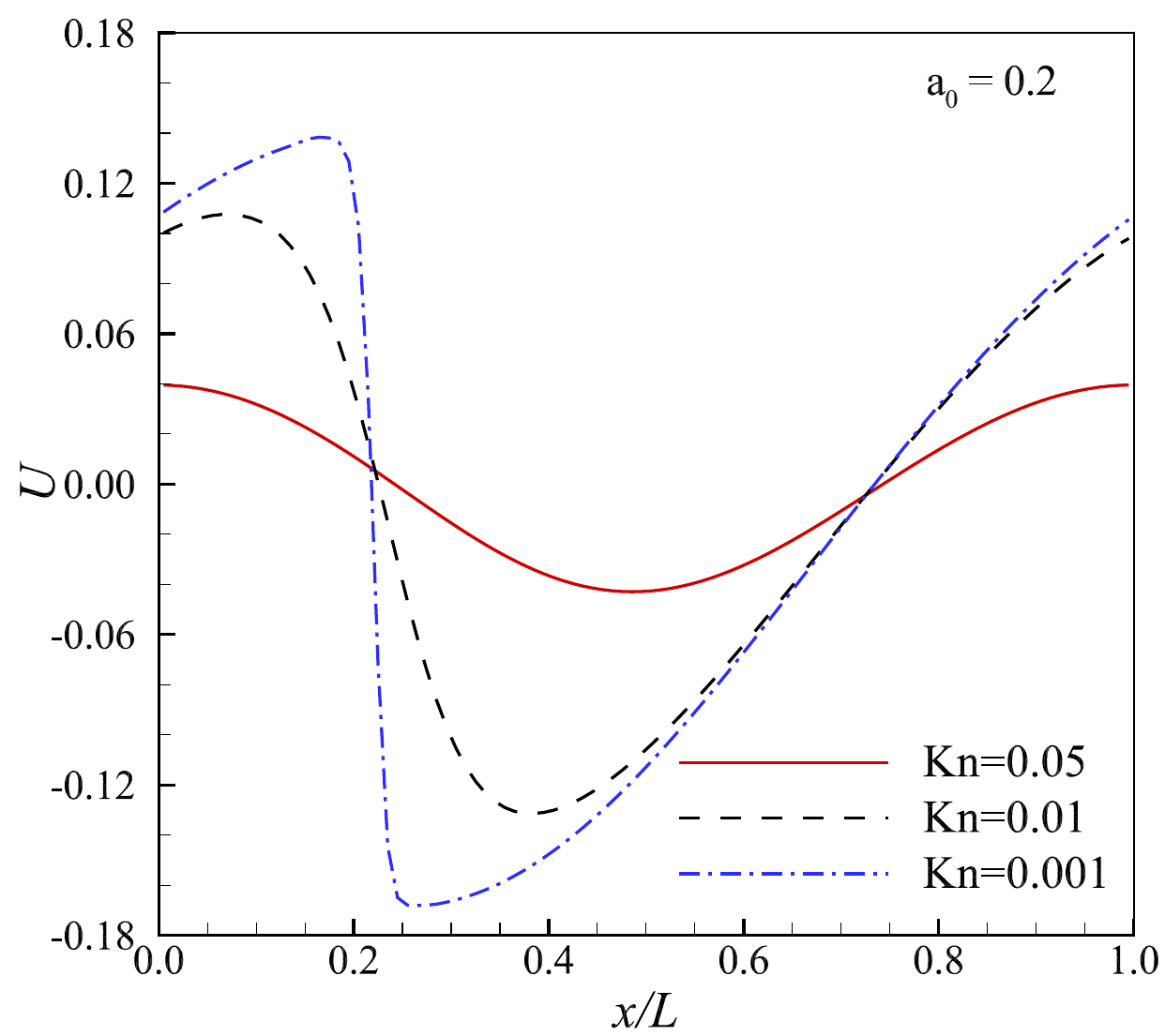}
    	}
	\caption{Density fluctuations and velocity distributions along the wavelength at different $\mathrm{Kn}$.}
    \label{fig: rho/u_x_a}
\end{figure}

Fig.\ref{fig: rho/u_x_a} presents the density and velocity distributions along the wavelength at the end of a single period after the calculation converges, from $a_0 = 0.01$ to $a_0 = 0.2$. As shown in the figure, the distribution of density exhibits a shape similar to that of a cosine function, which is consistent with the shape of the acceleration, and meanwhile, the shape of the velocity distribution is analogous to that of the number density distribution—a pattern that stems from two key interconnected effects. From the perspective of the macroscopic equations (as presented in Eq.\ref{eq: Acceleration term 1 macro}), the source term acts on the momentum. Therefore, the convection term drives density variations and induces shifts in the distribution of all macroscopic variables. Specifically, the higher the macroscopic velocity, the more pronounced this deviation of the number density fluctuation distribution from the cosine function becomes. As can be seen from Fig.\ref{fig: rho_x_a0.01} to\ref{fig: rho_x_a0.2}, when $\mathrm{Kn}=0.01$, the trough of the density fluctuation curve shifts from the position of 0.475 at $a_0 = 0.01$ to 0.385 at $a_0 = 0.2$, a displacement that results in significant deformation of the density distribution curve for $a_0 = 0.2$. Furthermore, it can be observed from the figures that when $Kn = 0.05$, the trough displacement deviation is smaller, primarily due to reduced relaxation effects. As exemplified by the free molecular flow regime, the macroscopic velocity no longer influences variations in the microscopic distribution function under such conditions. In contrast, for $\mathrm{Kn} = 0.001$, the larger deviations in both amplitude and displacement confirm that these deviations are convection term dominated, while viscosity exerts a suppressive effect on them, which represents the macroscopic manifestation of microscopic relaxation. Collectively, these cases quantitatively demonstrate the role of viscosity in mitigating such deviations.


\section{Conclusions}\label{sec:conclusion}

In this study, the Unified Gas-Kinetic Scheme (UGKS) was employed to simulate Coherent Rayleigh–Brillouin Scattering (CRBS), with the governing equation derived from the BGK–Shakhov model. To accommodate the additional perturbation source term in the governing equations, the Strang splitting method was adopted. The proposed model was validated through grid-independence tests and comparisons with argon experimental data, confirming its multiscale capability and numerical accuracy. Subsequently, the influence of particle acceleration induced by optical dipole forces on CRBS spectra across varying Knudsen numbers (Kn) was systematically investigated.

The results show that, for a given acceleration, an increase in Kn—representing the transition from continuum to rarefied regimes—produces a distinct evolution in CRBS spectra: from two Brillouin peaks to a mixed Brillouin–Rayleigh profile, and ultimately to a single Rayleigh peak. Within the examined acceleration range, variations in the acceleration magnitude ($a_0$) affect the CRBS spectra only in near-continuum conditions. When $a_0$ exceeds a critical threshold, subtle changes in line shape occur, resulting in a spectrum similar to that at higher Kn, indicative of a Brillouin-to-Rayleigh transition. Moreover, this critical $a_0$ decreases as Kn diminishes. The observed spectral variations arise primarily from changes in the convection term, which modulates density perturbations and thereby alters the CRBS spectral response.

Overall, this work demonstrates the reliability and robustness of UGKS for CRBS simulations and elucidates the mechanisms by which flow rarefaction and optical-force-induced acceleration influence spectral characteristics. The findings broaden the scope of CFD-based CRBS modeling, extending its applicability to high-intensity optical and rarefied gas environments. Future efforts will focus on extending the methodology to CRBS spectra of polyatomic molecular gases.


\section*{Data availability}

The data that supports the findings of this study are available within the article and its supplementary material.


\section*{Declaration of competing interest}

The authors declare that they have no known competing financial interests or personal relationships that could have appeared to influence the work reported in this paper.


\section*{Acknowledgements}

The current research is supported by National Natural Key R$\&$D Program of China (Grant Nos. 2022YFA1004500), National Natural Science Foundation of China (12172316, 92371107), and Hong Kong research grant council (16301222, 16208324).

\bibliography{ugkscrbsref}

\end{document}